\documentclass[aps,pra,twocolumn,superscriptaddress,longbibliography]{revtex4-1}%
\usepackage{graphicx}
\usepackage{float}
\usepackage{dcolumn}
\usepackage{bm,color}
\usepackage{amsmath,amssymb,dsfont,amstext,amsfonts}
\usepackage{extarrows}
\usepackage[colorlinks=true,linkcolor=blue,urlcolor=blue,citecolor=blue]{hyperref}
\usepackage{xcolor}
\usepackage{wasysym}
\usepackage{mathtools}
\usepackage{bbold} 

\usepackage[normalem]{ulem} 
\usepackage{color}

\newcommand{\ra}{\rangle}
\renewcommand{\ao}{\hat{a}^{\phantom{\dag}}}
\renewcommand{\aa}{\hat{a}^\dag}
\newcommand{\Ho}{\hat{H}}

\newcommand{\mom}{\textbf{\textit{k}}}

\newcommand{\pauli}{\boldsymbol{\sigma}}

\begin{document}
\title{ Wave packet dynamics and edge transport in anomalous Floquet topological phases }
\author{Miguel F.~Mart\'{i}nez}
\affiliation{Department of Physics, KTH Royal Institute of Technology, 106 91, Stockholm, Sweden}
\affiliation{TCM Group, Cavendish Laboratory, University of Cambridge, JJ Thomson Avenue, Cambridge CB3 0HE, United Kingdom\looseness=-1}
\author{F.~Nur \"{U}nal}
\email{fnu20@cam.ac.uk}
\affiliation{TCM Group, Cavendish Laboratory, University of Cambridge, JJ Thomson Avenue, Cambridge CB3 0HE, United Kingdom\looseness=-1}

\begin{abstract}
The possibility of attaining chiral edge modes under periodic driving has spurred tremendous attention both theoretically and experimentally, especially in light of anomalous Floquet topological phases that feature vanishing Chern numbers unlike any static counterpart.
We here consider a periodically modulated honeycomb lattice and experimentally relevant driving protocols, which allows us to obtain edge modes of various character in a simple model. We calculate the phase diagram over a wide range of parameters and recover an anomalous topological phase with quasienergy gaps harbouring edge states with opposite chirality. 
Motivated by the advances in single-site control in optical lattices, we investigate wave packet dynamics localized at the edges in distinct Floquet topological regimes that cannot be achieved in equilibrium. 
We analyse transport properties in edge modes which originate from the same bands, but with support at different quasienergies and sublattices as well as possessing different chiralities. 
We find that an anomalous Floquet topological phase can in general generate more robust chiral edge motion than a Haldane phase, allowing for more effective loading of the wave packet into edge channels.
Our results demonstrate that the rich interplay of wave packet dynamics and topological edge states can serve as a versatile tool in ultracold quantum gases in optical lattices.
\end{abstract}

\maketitle

\section{Introduction}
Topologically protected phenomena entail a prominent research direction in condensed matter physics~\cite{Hasan_RevModPhys,Qi_RevModPhys}. A wide range of novel phases arising from the interplay of topology and symmetries have been theorised and observed, with intriguing features being unearthed regularly especially in highly nontrivial many-body or out-of-equilibrium settings~\cite{TKNN, bernevig_2006, Fang_2012, KaneMele_Z2,  Slager13_NatPhys, Kruthoff_2017, PoVishwanath17_NatComm, Bradlyn17_Nat, Jotzu14_Nat,Aidelsburger15_NatPhys,Tran17_SciAdv_dichrosim,Asteria19_NatPhys,Kemp22_PRR,TanYu19_PRL_QGTmeasurement, Zhai2022,Vahid2022, jangjan2020}. Regarding the latter, rapid developments have extended topological characterisations to periodically driven Floquet systems~\cite{Roy_PRB, Oka_PRB,Kit_PRB,Rudner_PRX,Nur_PRL_HowToMeasure,wintersperger2020realization} and dynamic quench settings with new invariants~\cite{Wang_2017,Tarnowski19_NatCom,Unal19_PRR_hopf,HuZhao20_PRL_hopf}, even reaching to exotic multi-gap topologies with non-Abelian braiding properties~\cite{Unal_2020,Slager22_arXivAnomEuFloq, Ahn2019, bouhon2019nonabelian,bouhonGeometric2020,Jiang2021,Jiang1Dexp}. 
From an experimental point of view, Floquet engineering~\cite{Goldman_PRX,Eckardt_RevModPhys,Bukov_AdvPhys} has been established as a powerful tool to realise paradigmatic models in periodically driven non-equilibrium quantum matter in platforms such as ultracold atoms~\cite{Cooper19_RMP,WangUnal_18_PRL,RaciunasUnal_18_PRA,wintersperger2020realization, Reichl14_PRA} and photonic lattices~\cite{Maczewsky17_NatCommun,Mukherjee17_NatComm}, allowing for not only high degree of control and efficient quantum simulations but also exploring new regimes unattainable in equilibrium.


%
In a Floquet system, where energy is not a conserved quantity due to broken continuous time-translation invariance, one can adopt a description in terms of a periodic quasienergy since discrete time translations are still present. An effective quasienergy spectrum and the topological information are encoded by the time evolution over a period, $T$. Upon evaluating stroboscopically~\cite{Eckardt_RevModPhys}, a quasienergy can be defined as phase eigenvalues of the time evolution operator, namely as $\varepsilon_nT \in [-\pi,\pi)$ for $n$ number of bands, modulo $2\pi$ in a Floquet Brillouin Zone (FBZ). 
The fact that quasienergy bands are phases forming a circle induces one additional, {\it anomalous}, $\pi$-gap connecting the bands through the FBZ edge. The periodicity of the Floquet spectrum has paved the way for novel phases that truly arise from this out-of-equilibrium nature such as helical edge states crossing across the FBZ, anomalous Floquet Anderson insulators and anomalous Dirac string phases~\cite{Budich17_PRL_helical,Titum16_PRX,Slager22_arXivAnomEuFloq}.

Most interestingly, the possibility to obtain anomalous edge states in the FBZ-edge gap renders the equilibrium topological classification in terms of the Chern number, $C_n$, in two dimensions inept to characterize driven systems~\cite{Kit_PRB,Rudner_PRX}. Rather than invariants of individual bands, one needs to consider winding numbers, $W_{g}$, associated with gaps centered around, e.g.~$g=0$ and $g=\pi$ for two levels. Consequently, the anomalous Floquet topological phase has attracted great attention, characterised by the winding number combination $[W_0, W_\pi]=[1,1]$ harbouring edge states in both gaps despite a vanishing Chern number~\cite{wintersperger2020realization, Rudner_PRX}. Experimentally, individual edge modes have been probed in photonic lattices~\cite{Maczewsky17_NatCommun,Mukherjee17_NatComm}. Advances in optical lattices have allowed for directly measuring the topological invariants~\cite{wintersperger2020realization,Tarnowski19_NatCom,Jotzu14_Nat} and in particular distinguishing the quasienergy gaps to unambiguously assign the observed winding numbers to individual gaps. 
However, coherent edge dynamics and transport properties in different quasienergy gaps, particularly with respect to each other and equilibrium phases, remain an open question. Recent advances in single-site accessibility~\cite{GrossBakr21_NatPhys} in optical lattices now offer new possibilities for the creation of sharp edges and probing topological edge modes by using localised wave packets to investigate unique Floquet topological features as well as the effect of different quasienergy gaps associated to different branch cuts. 


In this work, we consider a periodically driven two-band model in two dimensions (2D) that is also experimentally relevant and analyse transport properties in different quasienergy gaps focusing on the distinct Floquet nature~\cite{Nur_PRL_HowToMeasure,wintersperger2020realization}. We calculate the phase diagram over a wide range of parameters and contrast different driving protocols. Going beyond the anomalous $[1,1]$-phase that has been originally introduced, this allows us to reach an unexplored anomalous Floquet topological phase where a pair of edge states with opposite chiralities are induced in different gaps ($[W_0, W_\pi]=[\pm 1, \mp 1]$) supported by the same two bands with finite Chern number, which cannot be obtained in equilibrium. We investigate the wave packet dynamics in various topological, and in particular anomalous, phases. 
We study populating edge modes at different quasienergies and with different winding numbers by applying kicks, controlling the shape of the wave packet, and examine their robustness and efficiency of preparation. Addressing chiral edge dynamics in different phases where only a single gap or both gaps harbour edge modes, we show that an anomalous Floquet topological phase can give rise to much more robust edge transport than equilibrium Chern insulating phases. We further analyze the effect of the Floquet gauge and the sublattice character of the edge states.

\vspace{-3mm}
\section{The Model}  \label{sec:Model}   
We consider a honeycomb lattice in two dimensions with a Hamiltonian given by
\begin{equation}
\label{eqn: hamiltonian general}
\Ho=-\sum_{\langle i,j \rangle} J \aa_i \ao_j+
\dfrac{\Delta}{2}\sum_{i\in A}\aa_i \ao_i -\frac{\Delta}{2}\sum_{i\in B}\aa_i \ao_i ,
\end{equation}
where $ \aa_i (\ao_i)$ creates (annihilates) a particle on lattice site $i$, with a nearest-neighbor tunnelling strength $J$ and an energy offset $\Delta$ between the two sublattices $A$ and $B$. We will introduce the periodic driving via the modulation of the hopping amplitudes $J_m$ for $m=1,2,3$ along the three nearest-neighbor vectors $\bm{d}_1=(0,-1)a,\,\bm{d}_2=(-1/2,\sqrt{3}/2)a$ and $\bm{d}_3=(1/2,\sqrt{3}/2)a$, where $a$ is the nearest-neighbor distance. The Hamiltonian at a given time instance is diagonal with respect to quasimomentum $\mom$, and hence can be written as
\begin{equation}
\label{eqn:momentum hamiltonian}
\Ho(\mom, t)= -\sum_{m=1}^3 J_m(t) \big(\cos(d_m \mom) \sigma_x+\sin(d_m \mom) \sigma_y \big) +\frac{\Delta}{2}\sigma_z ,
\end{equation}
where $\pauli$ are the Pauli matrices.

The driving protocols that we implement are of step-wise nature~\cite{Kit_PRB,Rudner_PRX}, which not only offers conceptual simplicity for our theoretical characterisation, but are also experimentally relevant as they have been recently implemented in optical lattices~\cite{wintersperger2020realization} with a smoothed modulation~\cite{Quelle17_NJP}. Namely, one period of the drive is divided in three even steps of length $T/3$. For the first driving scheme, the tunnelling is allowed cyclically only along one of the three directions during each stage with amplitude $J$~\cite{Rudner_PRX,Nur_PRL_HowToMeasure}. Secondly, we employ another protocol, where during the $m^{th}$-step, the tunneling $J_m=\lambda J$ is enhanced by a factor of $\lambda$, while the tunneling along the other two directions are kept fixed at $J$~\cite{Kit_PRB}. The first drive can be seen as a limiting case of the second one for $\lambda\rightarrow \infty, J\rightarrow0$, while keeping $\lambda J$ fixed. We will illustrate in detail the difference between the two schemes in the following.

Since during each step of the driving cycle, the Hamiltonian \eqref{eqn:momentum hamiltonian} becomes time-independent, $\Ho_m(\mom)$, for both driving protocols, the time-evolution operator at the end of one period can be written as
\begin{equation}
\label{eqn: floquet hamiltonian drive}
\hat{\mathcal{U}}(\mom,T)=e^{-i\Ho_3(\mom)\frac{T}{3}}e^{-i\Ho_2(\mom)\frac{T}{3}}e^{-i\Ho_1(\mom)\frac{T}{3}}=e^{-i\hat{\mathcal{H}}_F(\mom)T},
\end{equation}
where we set $\hbar=1$ throughout this paper. This stroboscopic evolution is captured by the Floquet Hamiltonian $\hat{\mathcal{H}}_F(\mom)$, defining the quasienergy spectrum through $\hat{\mathcal{H}}_F(\mom)|\phi_n(\mom)\ra=\varepsilon_n|\phi_n(\mom)\ra$. The Berry curvature and the Chern numbers of these quasienergy bands are calculated using the Floquet eigenstates $\phi_n(\mom)$. These topological invariants in the 2D momentum-space BZ have, however, been shown to be insufficient to capture the Floquet topology~\cite{Rudner_PRX}. One instead needs to consider also the time evolution within the period, $\hat{\mathcal{U}}(\mom,t)$.

\begin{figure}[t]
  \includegraphics[width=1\linewidth]{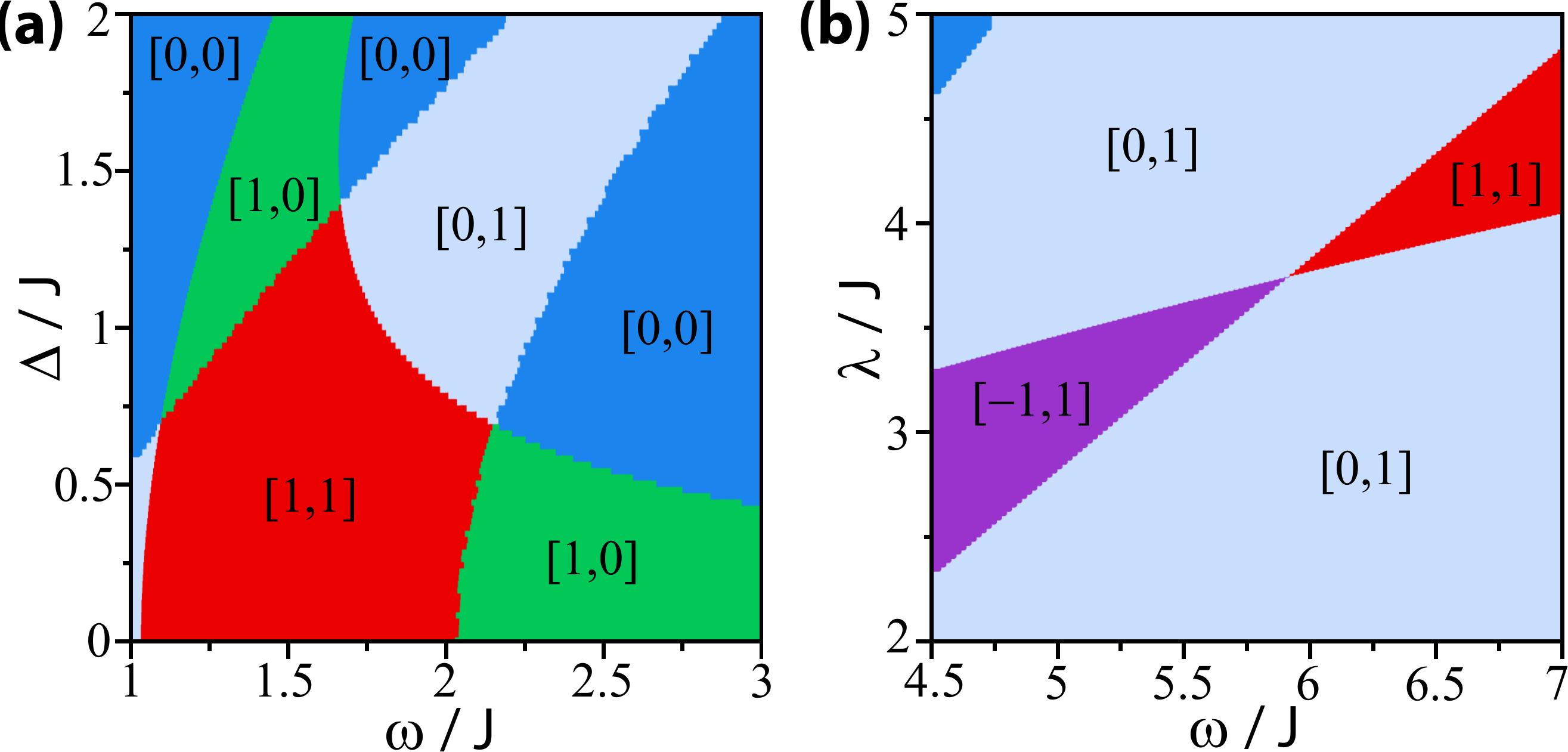}
    \caption{ Phase diagrams of the step-wise driven honeycomb lattice with corresponding winding numbers $[W_0,W_{\pi}]$.~(a) For the first driving protocol with switching tunnelling amplitudes along the three directions on/off completely, for modulation frequency $\omega$ and sublattice offset $\Delta$. (b) The second protocol where tunnelling amplitudes are cyclically enhanced by a factor of $\lambda$, here $\Delta=2J$. }
    \label{fig:PhaseDiagram}
\end{figure} 
In Floquet settings, the two bands can close and re-open in two distinct ways; in the quasienergy gaps at zero but also at $\pi$, corresponding to a change in the branch cut for defining $\hat{\mathcal{H}}_F(\mom)$, as opposed to one possibility in a static system. This quasienergy gap labelling originating from the time-periodicity requires a topological characterisation that employs winding numbers defined in the $(k_x,k_y,t)$-space. Each gap closing induces a transfer of Berry curvature between the bands, leaving chiral edge states behind in their respective gaps characterized by finite winding numbers. When the transitions in zero and $\pi$ gaps trivialise each other, we arrive at an anomalous Floquet topological phase with a vanishing Chern number~\cite{Rudner_PRX}. The latter can in general be expressed as the difference of the winding numbers (net number of edge states in a gap factoring in their chiralities) above and below a band, $C_n=W_{n,above}-W_{n,below}$. In the Floquet case, the extra $\pi$-gap renders the spectrum unbounded and, hence, offers more interesting possibilities. 

\vspace{-3mm}
\section{Phase diagrams}     
Fig.~\ref{fig:PhaseDiagram} demonstrates the phase diagrams that we numerically calculate in our driven honeycomb models for a representative parameter range. 
The winding numbers can in general be computed using the time evolution operator at every point in the $(2+1)$D parameter space~\cite{Rudner_PRX}, although this may prove computationally and experimentally demanding in most cases. Instead, we here employ an approach based on tracking the change of the winding numbers in each gap as introduced in Ref.~\cite{Nur_PRL_HowToMeasure} and successfully implemented in Munich~\cite{wintersperger2020realization} to measure anomalous winding numbers.
In particular, in the high-frequency regime we utilise the equilibrium topological classification based on Chern numbers with a trivial winding number in the $\pi$-gap~\cite{Nur_PRL_HowToMeasure,nathan2015topological}. In the case of the second driving protocol, $\lambda=1$ automatically satisfies the static definition. We depart from the topological invariants that we calculate for these initial parameters at high frequencies for both driving protocols. As model parameters are being tuned, we compute the winding numbers, which change via band touching points in each gap, by evaluating the charge of these topological singularities in a gap-specific way (see Supplementary Material for details~\cite{SeeSupplement}). The band singularities (hence, edge modes) at the FBZ edge involve a change in the branch cut by $\pi$. 
We further confirm these winding numbers by computing the Hopf invariant at representative points in a given phase, 
which has been shown to equal the winding numbers~\cite{Unal19_PRR_hopf}.

\begin{figure}
  \includegraphics[width=1\linewidth]{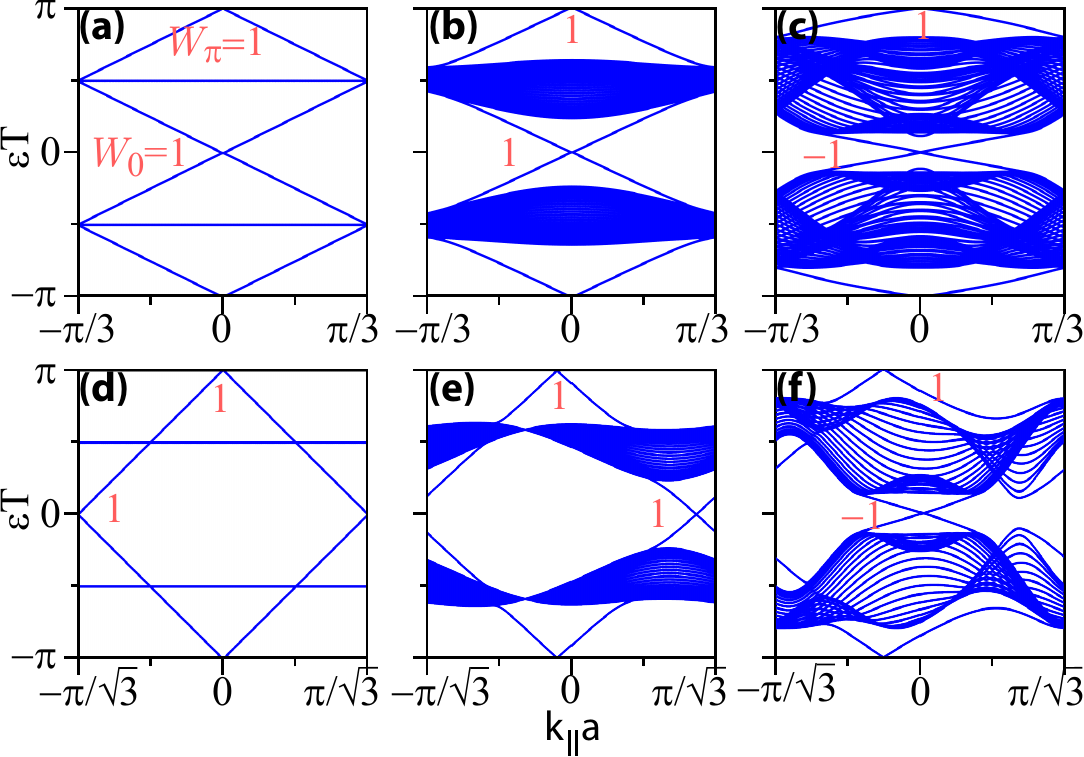}
   \caption{Quasienergy spectra for a ribbon with armchair (upper panel) and zigzag (lower panel) terminations, with crystal momentum $k_{\parallel}$ along the periodic direction. In the $[1,1]$ phase, here given for the first driving protocol introduced in the main text, zero- and $\pi$-gap edge states occur well-separate in momentum at the fine-tuned point $\omega=4J/3,\Delta=0$ (a,d) and $\omega=1.5J,\Delta=0.5J$ (b,e). (c,f) In the $[-1,1]$ phase with the second driving scheme, the two edge states appear closer in momentum, for $\omega =4.5J$, $\Delta=2J$, $\lambda=3$. 
   }
   \label{fig:Strip Band structures}
\end{figure}

For the first driving protocol where the tunneling amplitudes are cyclically turned on and off completely, the relevant tuning parameters are the sublattice offset and frequency. Indeed, this simple model illustrates a rich phase diagram including the previously predicted and observed anomalous Floquet topological phase $([1,1])$ (see Fig.~\ref{fig:PhaseDiagram}a), which can be understood by considering the limit of $\Delta=0$ and $\omega=4J/3$. For particles starting from one of the sublattices, this fine-tuned point corresponds to a complete population transfer to the other sublattice at the end of each step with tunnelling allowed for a time of $JT/3=\pi/2$. As we follow the driving cycle, it can be easily seen that particles in the bulk remain localized and only circularly move around each hexagon, ending up in alternating sublattice flavor at the end of each period and, hence, mixing the pseudospin character. However, in a finite system, 
particles move along the edge in a direction set by the chirality of the drive, corresponding to unit winding numbers of same sign in both gaps despite the trivial invariant of the bulk bands.

The second driving scheme on the other hand provides one more knob to tune, namely the driving amplitude $\lambda$. This allows for reaching more exotic phases as we present in Fig.~\ref{fig:PhaseDiagram}b as a function of $\lambda$ and $\omega$ for a fixed sublattice offset $\Delta=2J$. We identify a phase with winding numbers $[W_0,W_{\pi}]=[-1,1]$, which we numerically verify to be inaccessible using the first driving protocol owing to the less number of tuning parameters in that case. Distinct from the previously introduced anomalous phase, this is a hitherto-unexplored anomalous Floquet topological phase that harbours edge states in both zero and the anomalous--$\pi$ gaps with {\it opposite chiralities}, supported by the same two bands. Hence, the lower (upper) band carries a Chern number $C_1=-2 \; (C_2=+2)$. This anomalous phase is unique to the driven system, since in equilibrium there is only one gap where the topological transition can occur between the two bands (i.e.~the zero-gap). This gap could in principle host two edge modes of opposite chiralities also in the static case, provided that they occur at two different quasimomenta. This would however correspond to vanishing Chern numbers of the two bands, making the anomalous $[-1,1]$ phase exclusively emerging in the Floquet setting. Interestingly, these subtle differences also reflect on the edge transport and wave-packet dynamics as will be illustrated subsequently.

\section{Anomalous edge states}
In order to investigate transport properties and chiral edge dynamics in different Floquet (anomalous) topological phases, we consider a ribbon geometry extended along the $x$-direction, with $N_y$ layers along the finite $y$-direction. We present the edge spectra in Fig.~\ref{fig:Strip Band structures} for both armchair and zigzag terminations. At the fine-tuned point within the $[1,1]$ phase of the first driving protocol, the Floquet spectrum features completely flat bands corresponding to the localised bulk motion with extended edge modes crossing the entire FBZ, see Fig.~\ref{fig:Strip Band structures}(a,d). Although the armchair termination folds these two edge states to the same point, the zigzag spectrum reveals that the edge modes are well separated in momentum: While the zero-gap states form at the $K$-point with finite momentum $\pi/\sqrt{3}$ in units of the lattice constant, $\pi$-gap states appear at the $\Gamma$-point. We find that this is true in general in the $[1,1]$ phase owing to the nature of band inversions required in this phase, also away from the fine-tuned case as illustrated in Fig.~\ref{fig:Strip Band structures}(b,e) with dispersive bands, as well as for the second driving protocol. 
Since armchair and zigzag terminations correspond to projecting along perpendicular directions in the momentum space, $k\rightarrow-k$ symmetry is naturally broken in the presence of a finite sublattice offset for the latter (see Fig.~\ref{fig:Strip Band structures}(b,e)). The edge modes nonetheless still carry a large momentum difference.
On the contrary, in the $[-1,1]$ phase in Fig.~\ref{fig:Strip Band structures}(c,f), the two edge modes with opposite chiralities appear closer in momentum, which will bring about an important distinction for the dynamics in the two anomalous phases.


\begin{figure}
  \includegraphics[width=1\linewidth]{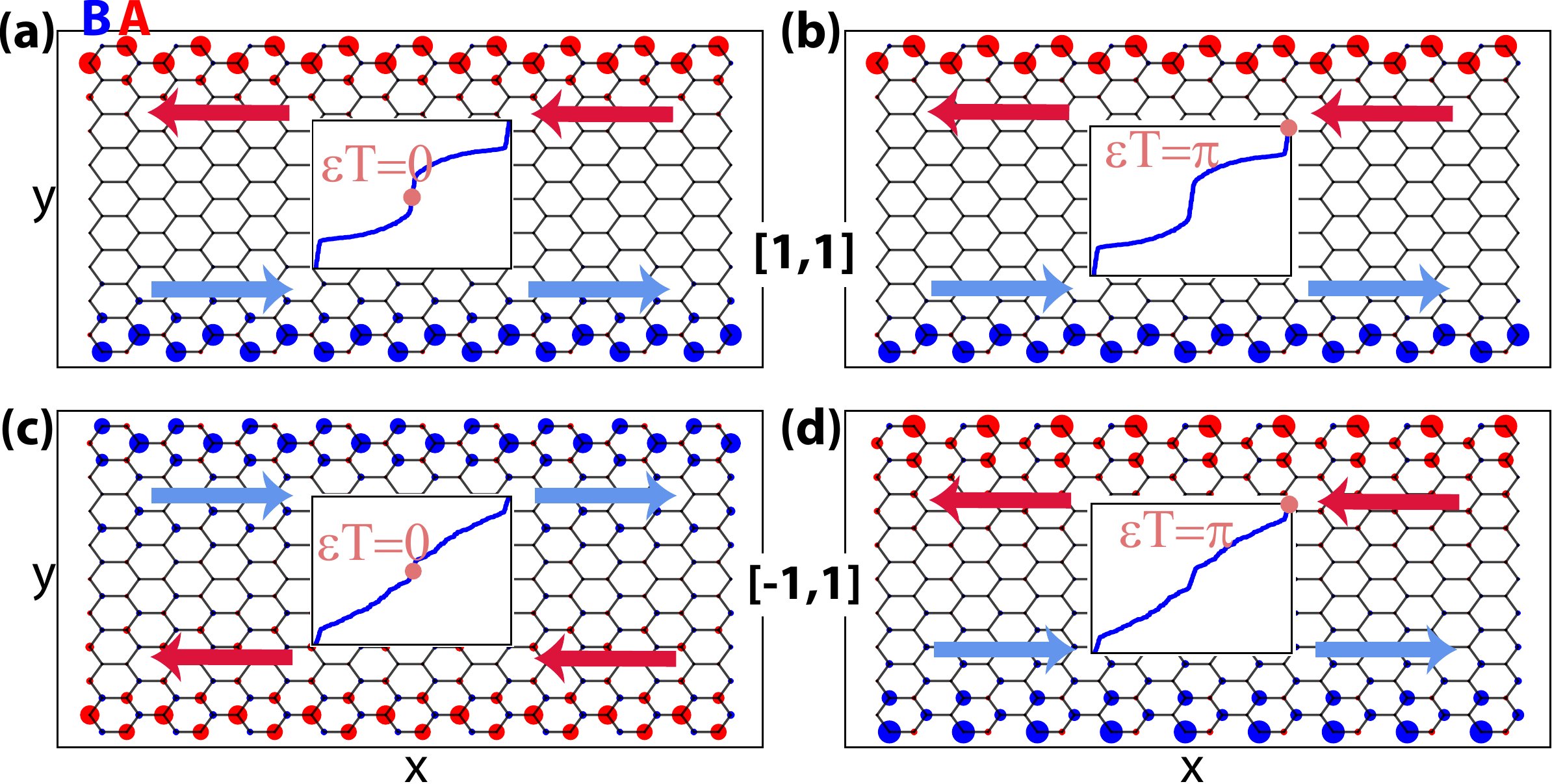}
    \caption{Distribution of edge states at quasienergy zero (left panel) and $\pi$ (right panel), marked by dots in the insets on their corresponding quasienergy spectra, on a cylinder periodic along the $x$-direction. (a,b) In the $[1,1]$ phase with the same parameters as Fig.~\ref{fig:Strip Band structures}b, both edge states are counterclockwise and localised at the same sublattices. (c,d) In the $[-1,1]$ phase for the parameters given in Fig.~\ref{fig:Strip Band structures}c, while the $\pi$-gap state localises on $A$ at the top edge, the zero energy state with the opposite chirality localises on the $B$ sublattice. 
    }
    \label{fig:EdgeStLocalisation}
\end{figure}

The different chiralities of the edge states in the anomalous Floquet topological phases also affect their sublattice character as shown in Fig.~\ref{fig:EdgeStLocalisation}. We here consider a cylinder geometry with periodic boundary conditions connecting $N_x$ layers along the $x$-direction.
While in the $[1,1]$ phase, the counter-clockwise edge states in the zero-gap are localised on the $A(B)$ sublattice on the upper (lower) end of the cylinder, the $\pi$-gap states support the same chiral motion localised on the same sublattice flavors. In the case of $[W_0,W_{\pi}]=[-1,1]$, however, the system harbours both clockwise and counter-clockwise edge modes. Fig.~\ref{fig:EdgeStLocalisation}(c,d) demonstrates that this is facilitated by swapping of the sublattice character of the zero-gap states along with their chirality. Hence, on the upper/lower end of the cylinder, the two edge modes support opposite currents in different sublattices. Interestingly, the layers where zero and $\pi$-gap currents have maximum density, depicted by the size of the circles, are also different. We emphasize that these two chiral modes do not hybridize despite being on the same edge as they are well separated in quasienergy. We now analyse how the interplay between the edge modes located at different momentum, quasienergy, sublattices and with different chiralities affect their transport properties, especially with respect to each other and in different Floquet (anomalous) topological phases.

\begin{figure*}
     \includegraphics[width=1\linewidth]{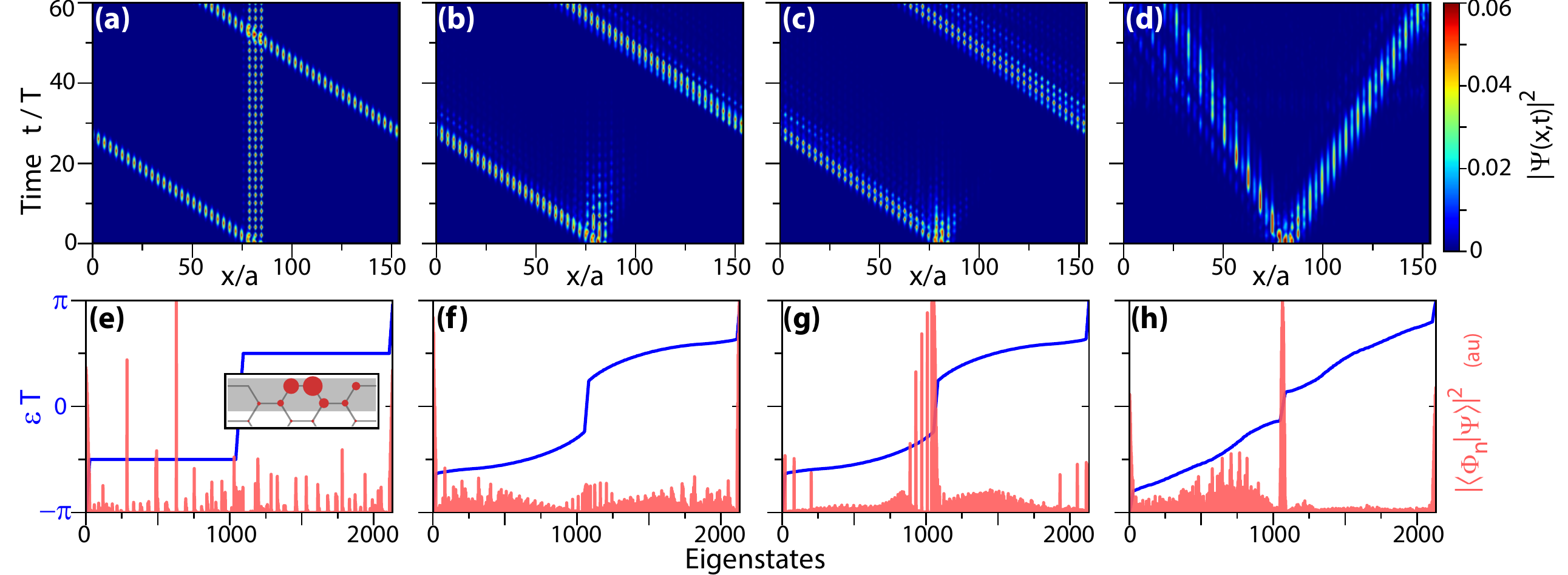}
    \caption{Overlap of a wave packet with the Floquet eigenstates (bottom panel) and its evolution (upper panel) at the edge sites (shaded two layers in the inset) on a cylinder periodic along $x$ with $N_x=104$, $N_y=41$ layers. The initial wave packets have $\sigma_x=1$, $\sigma_y=0.5$ as depicted in the inset where the radius of the circles is proportional to the initial probability at each site. (a, e) The $[1,1]$ phase at the fine-tuned point. The wave packet initialised without a kick follows a clear chiral motion. For the $[1,1]$ phase for the same parameters as in Fig.~\ref{fig:Strip Band structures}b, the wave packet is initially given (b,f) a small kick $\textbf{\textit{q}}=(-0.17, 0)$ and (c, g) a larger kick $\textbf{\textit{q}}=(1.56, 0)$, to target the Dirac cones at the $\pi$- and zero-gap respectively. As revealed by the overlaps, applying appropriate kicks allows us to mainly populate a given gap, where the dynamics evolve similarly. (d, h) The $[-1,1]$ phase for the same parameters as Fig.~\ref{fig:Strip Band structures}c. We now can largely populate both gaps simultaneously without a kick since the opposite chirality edge modes appear closer in momentum. The wave packet separates into two, giving rise to topologically protected currents travelling in both directions at the edge, with the Chern number ($|C|=2$) corresponding to the difference of them.}
    \label{fig:Anom.Dynamics}
\end{figure*} 

\section{Wave packet dynamics in anomalous topological phases}
Ultracold atomic systems have recently enjoyed rapid advances in single-site accessibility and local control with techniques like quantum gas microscopes and optical tweezers~\cite{GrossBakr21_NatPhys,KaufmanNi21_NatPhys,MunichExpWP,Nixon23_arxiv_localdrive}. Giving access to the creation of sharp edges, these pose a timely question and offer new opportunities for investigating wave packet dynamics localised at the edges of a topological system as a powerful tool in optical lattices.
The models that we implement allow us to retrieve a rich phase diagram within a simple geometry, where we compare edge transport in the conventional Haldane ($[1,0]$)~\cite{Haldane88_PRL} and the Haldane-like phases ($[0,1]$), which are gauge equivalent, with the anomalous Floquet topological phase ($[1,1]$). Indeed, we show that edge state population in a given gap can be mostly controlled by employing appropriate kicks. Most importantly, we find that the $[1,1]$ phase allows for a more robust chiral edge motion than the Haldane phases. We analyse the effects of opposite chiralities with the opportunity provided by the second anomalous phase ($[-1,1]$) and the simultaneous activation of both edge modes giving rise to interesting chiral edge dynamics.

We consider a cylindrical geometry of $N_x$ by $N_y$ layers and a Gaussian wave packet, $\Psi(x,y)=\exp{\left\{-(x-x_0)^2/4\sigma^2_x-(y-y_0)^2/4\sigma^2_y + iq_x x  + iq_y y\right\}}/\mathcal{N}$, initially localized at the upper edge (see Fig.~\ref{fig:Anom.Dynamics}e inset), at $(x_0,y_0)$ with a spread given by $(\sigma_x,\sigma_y)$ and normalization $\mathcal{N}$. We allow for an initial kick with momentum $\textbf{\textit{q}}=(q_x, q_y)$ which can be applied to control the overlap of the wave packet with edge states that are projected from the specified momenta onto the edge. We numerically calculate the evolution of the wave packet and present the probability at the edge sites in the upper two layers. 
In the $[1,1]$ phase, a wave packet without any kick ($\textbf{\textit{q}}=0$) predominantly overlaps with the $\pi$-gap states in Fig.~\ref{fig:Anom.Dynamics}(e) and (f), at and away from the fine-tuned point, since these edge states form at $\Gamma$ point (c.f.~Fig.~\ref{fig:Strip Band structures}). The corresponding wave packet dynamics display a clear chiral motion for long times around the edge of the cylinder which is periodic along the $x$-direction. While some of the probability naturally disperses into the bulk for the dispersive $[1,1]$ phase (see Fig.~\ref{fig:Anom.Dynamics}b), at the fine-tuned point the edge states are exclusively localised at $A$ sublattices. The probability at $B$ sites, hence, follows a chiral motion around each hexagon with completely flat bulk bands, giving rise to some probability dwelling around the initial position at all times in Fig.~\ref{fig:Anom.Dynamics}a.

Due to the large separation of the zero and $\pi$-gap states in the $[1,1]$ phase, we can populate the edge modes at the zero-gap by applying an initial kick of amount $|\textbf{\textit{q}}|=|K|$ to the wave packet. Fig.~\ref{fig:Anom.Dynamics}(c,g) displays a similar chiral motion with the not-kicked wave packet that is visibly indistinguishable, which is now mainly supported by the zero-gap states. We note that the kick direction (along the chiral edge mode or opposite to it) in fact does not matter as it only gives an overall phase, where the overlaps with eigenstates do not change and we observe the same dynamics. Moreover, since these wave packets are well localised in position space, we still obtain some spurious probability at the other gap, with or without a kick. This originates from the extensive overlap of the edge modes from the two gaps in position space (see Fig.~\ref{fig:EdgeStLocalisation}(a,b)), despite the fact that they are well separated in momentum. 
Nevertheless, in both cases, we can control the chiral motion to be carried predominantly by the target quasienergy-gap modes. 
This shows that although these edge modes form at different gaps and despite the difference in the branch cuts by $\pi$ while defining them, the wave packets can be controlled to populate mainly a given edge mode by applying appropriate kicks depending on their localisation in quasimomentum rather than quasienergy.
We present wider wave packets in the Supplementary Material where the population of a given gap can be further increased, which could be experimentally realised for example by scanning a wider region with the laser initialising the wave packets.


On the other hand, in the [-1,1] phase achievable with the second driving protocol, both zero and $\pi$ gaps harbour edge modes but their separation in momentum is less pronounced. Both edge modes can then be extensively populated with the same wave packet as demonstrated in Fig.~\ref{fig:Anom.Dynamics}h without applying any kick. Distinctively, since the winding numbers have opposite signs, we observe that the wave packet separates into two (see Fig.~\ref{fig:Anom.Dynamics}d), with a topologically protected current going in {\it both} directions at the edge of the cylinder, where the $\pi$-modes travel slightly faster in accordance with the spectrum (c.f.~Fig.~\ref{fig:Strip Band structures}(c,f)). We emphasise that in this phase, the quasienery bands carry a Chern number $|C|=2$, which is visible in the {\it difference} of the chiral currents at the edge of the system, rather than two topologically protected channels travelling along the same direction that would be expected in a static setting.

\begin{figure}
  \centering\includegraphics[width=1\linewidth]{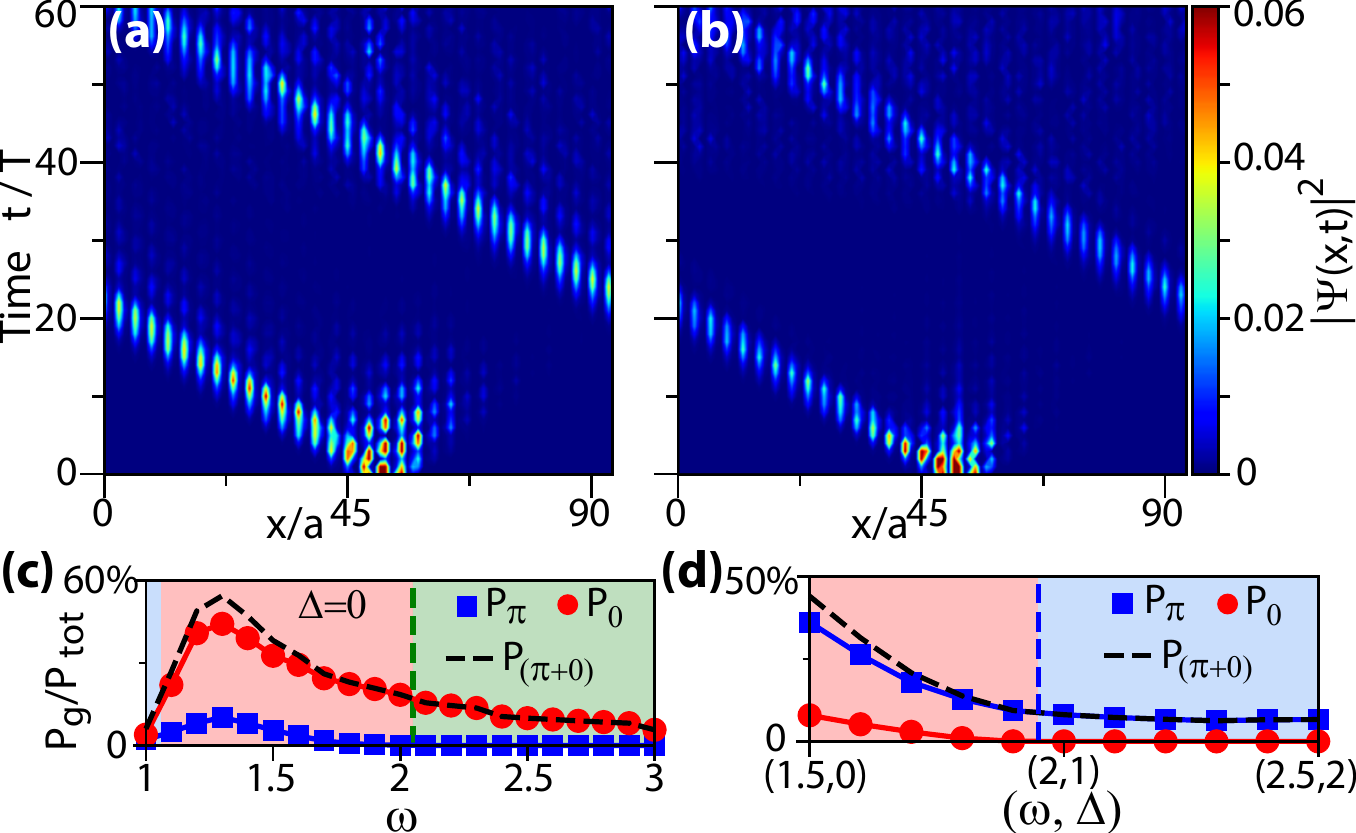}
    \caption{Edge dynamics of a wave packet (a) with an initial kick $\textbf{\textit{q}}=(\pi/\sqrt{3}, 0)$ in the $[1,0]$ phase for $\omega=2.2J$, $\Delta=0$, and (b) without a kick in the $[0,1]$ phase for $\omega=2J$, $\Delta=1J$. Both dynamics in the Haldane phases show qualitative differences with Fig.~\ref{fig:Anom.Dynamics}a-c, as they are less pronounced. (c) Percentage of the total probability carried by edge states in each gap, along the $\Delta=0$ cut of Fig.~\ref{fig:PhaseDiagram}a crossing different phases. The edge state population is much higher in the $[1,1]$ (red shaded area) than in the $[1,0]$ phase (green shaded area), for a wave packet with an initial kick as in (a) to target zero-gap states. (d) Similarly, the total probability per gap along a diagonal cut on the phase diagram crossing from $[1,1]$ to $[0,1]$, where the wave packet is given a small initial momentum to target the $\pi$-gap states forming close to $\Gamma$. A greater probability is supported by the two edge channels in the $[1,1]$ phase along both cuts (c,d), quantifying that edge transport is overall more robust in the anomalous phase. }
    \label{fig:HaldaneDynamics}
\end{figure}

The anomalous Floquet topological phases carry distinct features arising from their out of equilibrium nature. First of all, both edge modes behave effectively as one single channel in the $[1,1]$ phase due to their similar chirality. It is, hence, instructive to explore whether this anomalous Floquet topological phase (with two edge states of the same chirality at different quasienergy gaps) gives rise to a different edge transport than the Haldane phases with a single chiral mode, i.e.~to contrast anomalous Floquet topological phases with the equilibrium Chern insulating phases. We demonstrate the wave packet dynamics in the $[1,0]$ and $[0,1]$ phases in Fig.~\ref{fig:HaldaneDynamics}a and b respectively. While we apply a kick ($\textbf{\textit{q}}=K$) to the wave packet to target the zero-gap states in $[1,0]$ as they form at the $K$ point, we generally obtain a chiral transport at the edge as expected in both phases.

Most importantly, upon comparing with the Haldane phases as visible in their color scales, we observe that the wave packet dynamics in the anomalous Floquet phase $[1,1]$ is much more robust, with the edge separating more from the bulk. To quantify this finding, we consider cuts on the phase diagram (Fig.~\ref{fig:PhaseDiagram}a) across different phases and evaluate the total overlap of a wave packet with the edge states in Fig.~\ref{fig:HaldaneDynamics}(c,d). Targeting the zero-gap states with a kick, we indeed find that the total percentage of the edge state population is overall much higher in the $[1,1]$ than in the $[1,0]$ phase, where the former comprises mostly zero-gap states that are further enhanced by the $\pi$-gap contribution (see Fig.~\ref{fig:HaldaneDynamics}c). Similarly, Fig.~\ref{fig:HaldaneDynamics}d shows that the wave packet, now initialised without a kick, overlaps with the $\pi$-gap states much more in $[1,1]$ than $[0,1]$, where the zero-gap states further strengthen edge dynamics in the former. We observe that this behaviour is in general present across the phase diagram and also for the second driving protocol.

Remarkably, this demonstrates that a wave packet can be prepared to populate the edge modes more easily and efficiently in the anomalous phase than in Haldane phases. The anomalous topological phase then supports a more pronounced chiral edge motion with less leakage into the bulk. 
The more robust edge transport in the anomalous Floquet phase stems from the fact that the edge of the system accommodate two different channels rather than one, increasing the relative weight of the edge channels compared to the bulk so that they better separate spatially.
Compared to the topological phases with equilibrium counterparts, there is one more gap available to harbour edge states in the anomalous Floquet topological phase, which renders stronger edge transport possible supported with more edge states in the driven system. This can also aid novel anomalous Floquet Anderson phases under disorder~\cite{Titum16_PRX}.
We note that this qualitative analysis naturally depends on system details such as the bulk gap width and the properties of the phase. Indeed, we also obtain that the total edge population by a wave packet is overall larger for the $[-1,1]$ phase than the Haldane-like phase (see Supplementary Material). Although the trend is less pronounced due to different chiralities and parameter dependencies, the effect is still clearly discernible when the anomalous phase harbours two edge channels.

\begin{figure}
    \includegraphics[width=1\linewidth]{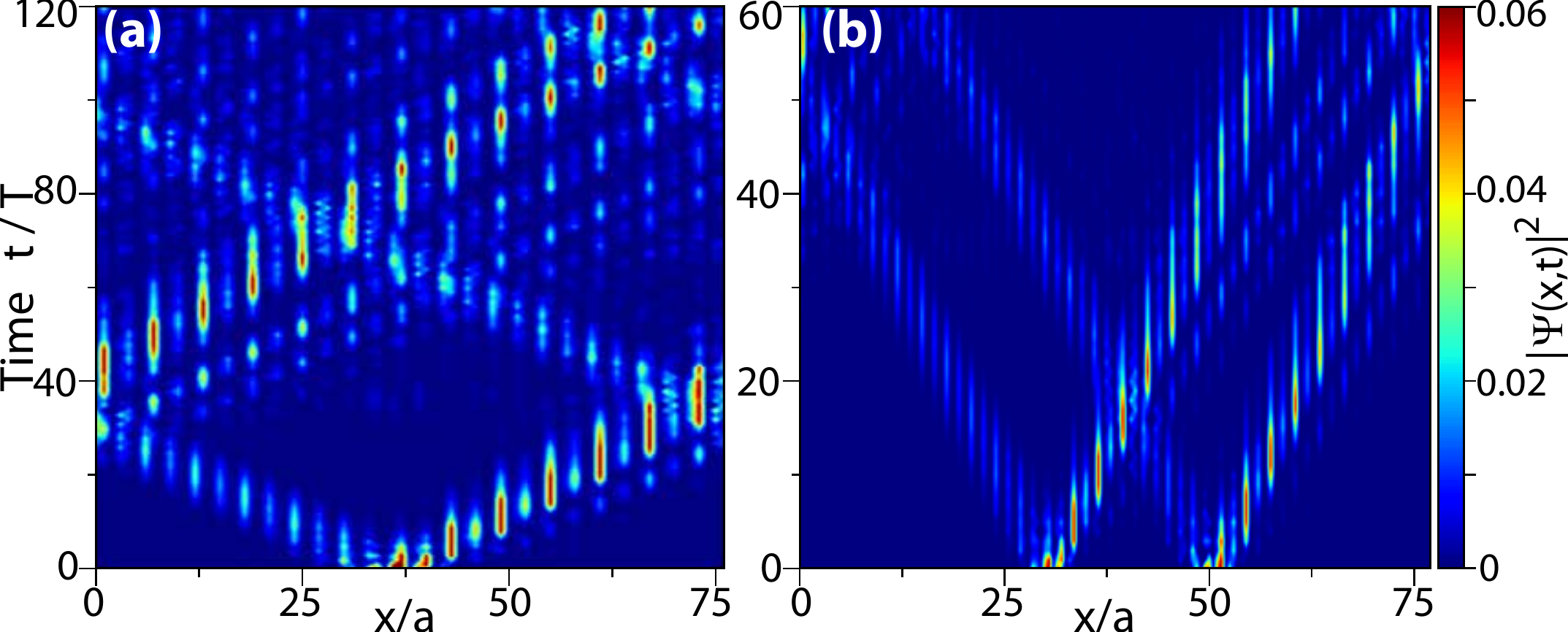}
    \caption{In the $[-1,1]$ phase with the parameters in Fig.~\ref{fig:Strip Band structures}c, the wave packet is initialised without a kick on a smaller cylinder with $N_x=52$, $N_y=41$. (a) Opposite going currents circle the cylinder and cross without disruption. (b) Two wave packets initialised at the edge pass through each other without hybridising as they are well separate in quasienergy. 
    }
    \label{fig:-1,1crossing} 
\end{figure}

The anomalous Floquet phase $[-1,1]$ manifests appealing features as well with the edge transport of opposite chiralities supported by the same bands. Since these edge modes are located at different quasienergy gaps and have support on different sublattices, they do not hybridize as demonstrated in Fig.~\ref{fig:-1,1crossing}a. When we consider a cylinder with a smaller size, we confirm that the opposite going currents at the edge circle the cylinder and pass through each other without disruption for several cycles. Similarly, when we introduce two wave packets, the crossing of the edge currents is clearly visible in Fig.~\ref{fig:-1,1crossing}b. We note that introducing two wave packets is an experimentally promising route for probing these anomalous topological dynamics with opposite chiralities, where localised particles can be created around circular-shaped sharp edges punched in a system with laser potentials.

Although the analysis carried out here is generic for any parameters in a given phase, the particular details of different edge currents depend on various aspects. For example, we observe some tails developing and travelling with different velocities in some dynamics. This generally arises when a lot of momentum states are stimulated due to the finite size of the wave packet, where the velocity of a given edge mode might be different at different quasimomenta (c.f.~Fig.~\ref{fig:Strip Band structures}) as well as different velocities that can be associated to different edge modes. We analyse some important details that can give rise to varying edge dynamics in a real system next.


\section{Floquet gauge and sublattice dependence}
For the wave packet dynamics presented in previous sections, we numerically calculate the time evolution by employing the Floquet Hamiltonian given in Eq.(\ref{eqn: floquet hamiltonian drive}). While such stroboscopic definitions are useful, we verify that these dynamic features are overall reproduced also by following the exact time evolution under the two driving protocols, justifying the employment of an effective description neglecting the micromotion~\cite{Eckardt_RevModPhys}. Although we here set the initial time to zero ($t_0=0$) for simplicity, the time evolution operator and hence the Floquet Hamiltonian in Eq.(\ref{eqn: floquet hamiltonian drive}) actually depend on this so-called Floquet gauge, $\hat{\mathcal{H}}^{t_0}_F(\mom)$, as well as the Floquet eigenstates~\cite{Eckardt_RevModPhys,Bukov_AdvPhys}. 

\begin{figure}
    \includegraphics[width=1\linewidth]{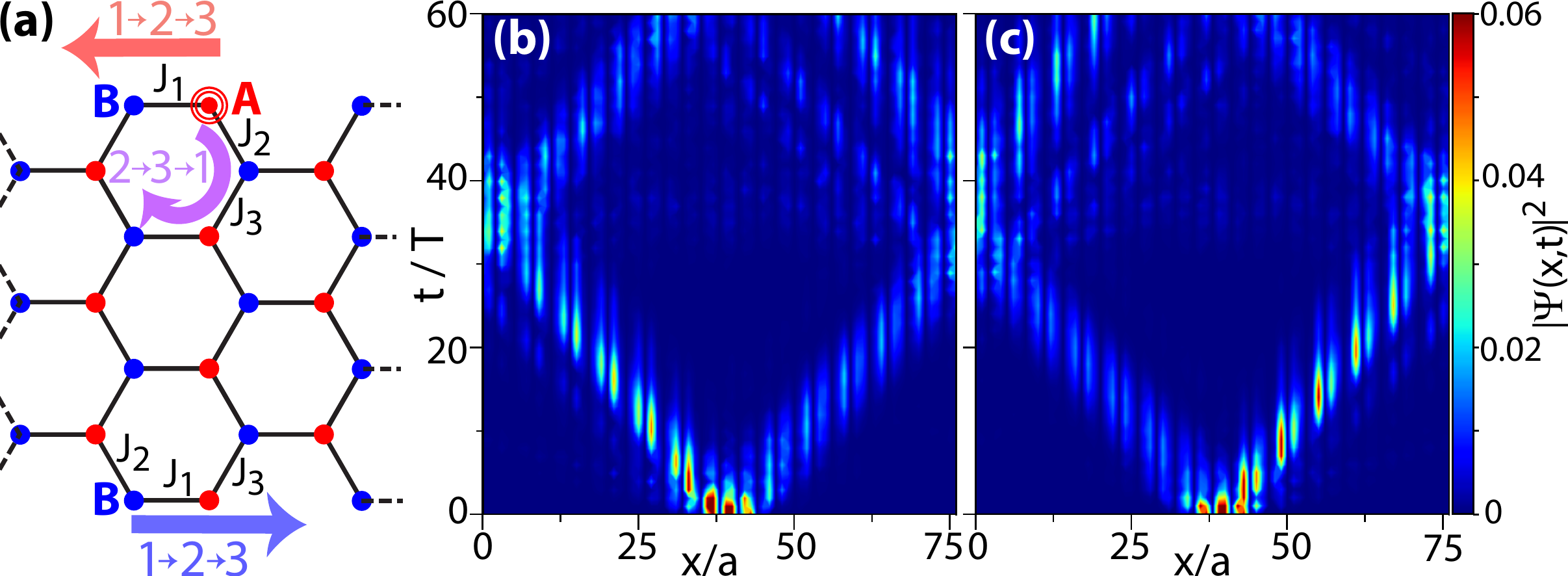}
    \caption{(a) Illustration of the relation between the upper/lower edge dynamics and the sublattice character by considering the fine-tuned point of the first driving protocol. Starting from $t_0=0$ and following the chiral motion with $J_1\rightarrow J_2\rightarrow J_3$ turned on cyclically, a particle localised on $A$ sublattice at the upper edge moves along the edge leftwards (red arrow). Starting from a different initial time of $t_0=T/3$ induces a bulk motion around the hexagon (purple). At the lower edge, the sublattice characters giving rise to the edge and bulk motions are swapped, where starting from $B$ at $t_0=0$ generates edge motion (blue). (b,c) Dynamics on a cylinder in the $[-1,1]$ phase for the same parameters as in Fig.~\ref{fig:Strip Band structures}c, where we localise the wave packet without a kick at the same $x$ position on the upper and lower edges respectively. We average over $20$ equally spaced initial times $t_0$ within a period, which produce similar chiral motion.}
    \label{fig:sublattice,AvgDyn}
\end{figure}

The Floquet gauge can influence the wave packet dynamics in experiments especially at lower frequencies where the details within a period become more crucial. We elucidate this by considering the extreme limits of a wave packet completely localised on an $A$ site at the upper edge in Fig.~\ref{fig:sublattice,AvgDyn}a and the fined-tuned point in the first driving protocol, where we obtain complete population transfer between sublattices at the end of each step of the drive. When we start from $t_0=0$ with the tunneling amplitudes are turned on/off cyclically as $J_1\rightarrow J_2\rightarrow J_3$, particles move along the edge leftwards. However, if we consider another initial time e.g.~$t_0=T/3$ within the same sequence (i.e.~$J_2\rightarrow J_3\rightarrow J_1$), particles would follow a chiral bulk motion around the hexagon. Away from the fine-tuned point, eigenstates mix sublattice flavors at varying degrees, but the Floquet gauge dependence remains.
While this can give rise to some modifications in exact dynamics observed, it does not alter our general conclusions. Furthermore, this is naturally less pronounced for more balanced or wider wave packets.

The upper and lower edges of the system may give rise to different dynamics as well, since the pseudospin character reflects onto these two physical edges oppositely. Namely, the chiral edge motion observed at the upper edge when we start from the sublattice $A$ as depicted in Fig.~\ref{fig:sublattice,AvgDyn}a can be obtained at the bottom edge starting from the $B$ sublattice (and similarly for the bulk localised states). When $\Delta=0$, the Hamiltonian can be cast into a diagonal form in the pseudospin. The up spin in the upper edge correspond to the down spin at the lower edge, which hence can bring about different dynamics depending on their population by the wave packet. While in general for $\Delta\neq0$, the spin characters are mixed, they are mixed in the same but opposite way at the upper and lower edges, i.e.~some ``up-down" mixture on one end corresponds to the opposite ``down-up" mixture on the other end. Changing the sign of the offset $\Delta\rightarrow-\Delta$ then inverts the upper/lower edge character, which we confirm in our numerics. 

Overall, when we average out over various dynamics, the upper and lower ends of the system naturally give rise to the same behaviour as it samples through different initial times and sublattice characters. We consider a wave packet localised at the same way at the upper and lower edges of a cylinder, at the same $x$ positions in the geometry depicted in Fig.~\ref{fig:sublattice,AvgDyn}a. We numerically calculate the chiral edge currents starting from different $t_0$ initial times (for 20 data points equally spaced over one period) which correspond to different eigenstates. We demonstrate the average wave packet dynamics at the respective edges in Fig.~\ref{fig:sublattice,AvgDyn}(b,c). While the upper/lower edge dynamics might be very different at a given $t_0$, we observe that the $\pi$-gap states generate the similar leftward/rightward currents at the top/bottom edges on average as expected, as well as the zero-gap modes supporting opposite chirality. We note that the negligible difference between the upper/lower dynamics in the figure stems from the coarse sampling for averaging, where more data points wash out these differences further.

\section{Conclusion}
In this work, we focus on the topological edge transport and chiral motion at the edge of a periodically driven system, particularly in anomalous Floquet topological phases unique to these out of equilibrium settings~\cite{Kit_PRB,Rudner_PRX}. In light of recent developments in single-site accessibility in optical lattices, which offers the possibility to probe the dynamics locally at sharp edges, we investigate wave packet dynamics as a versatile tool in exploring topologically protected Floquet edge modes. We consider a honeycomb lattice under experimentally relevant and conceptually insightful step-wise modulated driving protocols~\cite{Nur_PRL_HowToMeasure,wintersperger2020realization}, which allow us to retrieve a rich phase diagram, involving the conventional Haldane (-like) phases that can be realised in equilibrium as well as two different anomalous Floquet topological phases with no static counterparts that harbour edge states of different chiralities in both quasienergy gaps. 

We show that, in the $[1,1]$ phase, the edge modes living in different quasienergies behave predominantly like a single channel with the same chirality and sublattice character, where wave packets can be controlled to mostly populate one of the gaps by applying appropriate kicks as they are well separate in momentum. While this behaviour is similar to the Haldane phases with a topologically protected edge mode present only in a single gap~\cite{Haldane88_PRL}, we find that the anomalous $[1,1]$ phase can generally support a more robust edge transport with edge states separating more from the bulk. We explain this with having two different channels coming from both gaps accommodated at the edge of the system, which enhances the population of the edge modes by the wave packets, providing an overall advantage over Haldane phases with equilibrium counterparts.

We further show a hitherto-unexplored anomalous phase with opposite winding numbers ($[-1,1]$) that can be achieved using the second driving protocol in a honeycomb lattice, where the system features both clockwise and anti-clockwise currents now carried mainly by different sublattices at a given edge. Since the edge modes of opposite chiralities form much closer in momentum, we observe that wave packets localised at the edge largely populate both channels, which generates two independent currents going in opposite ways that can transverse through one another without hybridising. We also analyse the dependence of the dynamics on the details of the drive and the exact position where the wave packets are localised. We discuss that these effects can be controlled by changing the shape of the wave packet or averaging over different initial times. Our results demonstrate that investigating Floquet topological edge modes by using wave packets in optical lattices can reveal unique out-of-equilibrium features. These insights can be also employed in phases involving more number of edge states and with different chirality combinations~\cite{ZhangPan22_arXiv_floeuetBIS}, which might reveal more interesting edge dynamics and more robust edge transport engaging multiple edge channels. Photonic lattices employing quantum walks offer another promising route to study various anomalous Floquet phases with different winding number combinations~\cite{AdiyatullinAmo23_PRL}.

\textit{Note added}--We have became aware of a relevant experimental work during the completion of this manuscript, which has been announced following our work in Ref.~\cite{MunichExpWP}, studying and probing wave packet dynamics in an anomalous Floquet topological phase.

Data sharing not applicable to this article as no datasets were generated or analysed during the current study.

\begin{acknowledgments}
We thank Nigel Cooper and Robert-Jan Slager, for insightful discussions and comments on the manuscript, as well as the experimental team Monika Aidelsburger, Rapha\"el Saint-Jalm, Christoph Braun, Alexander Hesse and Johannes Arceri.
F.N.\"U. acknowledges funding from the Royal Society under a Newton International Fellowship, the Marie Sk{\l}odowska-Curie programme of the European Commission Grant No 893915, and Trinity College Cambridge. This work received funding from the European Research Council (ERC) under the European Union’s Horizon 2020 research and innovation program (Grant Agreement No. 101001902).
\end{acknowledgments}

\bibliography{references}

\begin{thebibliography}{57}%
\makeatletter
\providecommand \@ifxundefined [1]{%
 \@ifx{#1\undefined}
}%
\providecommand \@ifnum [1]{%
 \ifnum #1\expandafter \@firstoftwo
 \else \expandafter \@secondoftwo
 \fi
}%
\providecommand \@ifx [1]{%
 \ifx #1\expandafter \@firstoftwo
 \else \expandafter \@secondoftwo
 \fi
}%
\providecommand \natexlab [1]{#1}%
\providecommand \enquote  [1]{``#1''}%
\providecommand \bibnamefont  [1]{#1}%
\providecommand \bibfnamefont [1]{#1}%
\providecommand \citenamefont [1]{#1}%
\providecommand \href@noop [0]{\@secondoftwo}%
\providecommand \href [0]{\begingroup \@sanitize@url \@href}%
\providecommand \@href[1]{\@@startlink{#1}\@@href}%
\providecommand \@@href[1]{\endgroup#1\@@endlink}%
\providecommand \@sanitize@url [0]{\catcode `\\12\catcode `\$12\catcode
  `\&12\catcode `\#12\catcode `\^12\catcode `\_12\catcode `\%12\relax}%
\providecommand \@@startlink[1]{}%
\providecommand \@@endlink[0]{}%
\providecommand \url  [0]{\begingroup\@sanitize@url \@url }%
\providecommand \@url [1]{\endgroup\@href {#1}{\urlprefix }}%
\providecommand \urlprefix  [0]{URL }%
\providecommand \Eprint [0]{\href }%
\providecommand \doibase [0]{http://dx.doi.org/}%
\providecommand \selectlanguage [0]{\@gobble}%
\providecommand \bibinfo  [0]{\@secondoftwo}%
\providecommand \bibfield  [0]{\@secondoftwo}%
\providecommand \translation [1]{[#1]}%
\providecommand \BibitemOpen [0]{}%
\providecommand \bibitemStop [0]{}%
\providecommand \bibitemNoStop [0]{.\EOS\space}%
\providecommand \EOS [0]{\spacefactor3000\relax}%
\providecommand \BibitemShut  [1]{\csname bibitem#1\endcsname}%
\let\auto@bib@innerbib\@empty
\bibitem [{\citenamefont {Hasan}\ and\ \citenamefont
  {Kane}(2010)}]{Hasan_RevModPhys}%
  \BibitemOpen
  \bibfield  {author} {\bibinfo {author} {\bibfnamefont {M.~Z.}\ \bibnamefont
  {Hasan}}\ and\ \bibinfo {author} {\bibfnamefont {C.~L.}\ \bibnamefont
  {Kane}},\ }\bibfield  {title} {\enquote {\bibinfo {title} {Colloquium:
  Topological insulators},}\ }\href {\doibase 10.1103/RevModPhys.82.3045}
  {\bibfield  {journal} {\bibinfo  {journal} {Rev. Mod. Phys.}\ }\textbf
  {\bibinfo {volume} {82}},\ \bibinfo {pages} {3045--3067} (\bibinfo {year}
  {2010})}\BibitemShut {NoStop}%
\bibitem [{\citenamefont {Qi}\ and\ \citenamefont
  {Zhang}(2011)}]{Qi_RevModPhys}%
  \BibitemOpen
  \bibfield  {author} {\bibinfo {author} {\bibfnamefont {Xiao-Liang}\
  \bibnamefont {Qi}}\ and\ \bibinfo {author} {\bibfnamefont {Shou-Cheng}\
  \bibnamefont {Zhang}},\ }\bibfield  {title} {\enquote {\bibinfo {title}
  {Topological insulators and superconductors},}\ }\href {\doibase
  10.1103/RevModPhys.83.1057} {\bibfield  {journal} {\bibinfo  {journal} {Rev.
  Mod. Phys.}\ }\textbf {\bibinfo {volume} {83}},\ \bibinfo {pages}
  {1057--1110} (\bibinfo {year} {2011})}\BibitemShut {NoStop}%
\bibitem [{\citenamefont {Thouless}\ \emph {et~al.}(1982)\citenamefont
  {Thouless}, \citenamefont {Kohmoto}, \citenamefont {Nightingale},\ and\
  \citenamefont {den Nijs}}]{TKNN}%
  \BibitemOpen
  \bibfield  {author} {\bibinfo {author} {\bibfnamefont {D.~J.}\ \bibnamefont
  {Thouless}}, \bibinfo {author} {\bibfnamefont {M.}~\bibnamefont {Kohmoto}},
  \bibinfo {author} {\bibfnamefont {M.~P.}\ \bibnamefont {Nightingale}}, \ and\
  \bibinfo {author} {\bibfnamefont {M.}~\bibnamefont {den Nijs}},\ }\bibfield
  {title} {\enquote {\bibinfo {title} {Quantized hall conductance in a
  two-dimensional periodic potential},}\ }\href {\doibase
  10.1103/PhysRevLett.49.405} {\bibfield  {journal} {\bibinfo  {journal} {Phys.
  Rev. Lett.}\ }\textbf {\bibinfo {volume} {49}},\ \bibinfo {pages} {405--408}
  (\bibinfo {year} {1982})}\BibitemShut {NoStop}%
\bibitem [{\citenamefont {Bernevig}\ \emph {et~al.}(2006)\citenamefont
  {Bernevig}, \citenamefont {Hughes},\ and\ \citenamefont
  {Zhang}}]{bernevig_2006}%
  \BibitemOpen
  \bibfield  {author} {\bibinfo {author} {\bibfnamefont {B~Andrei}\
  \bibnamefont {Bernevig}}, \bibinfo {author} {\bibfnamefont {Taylor~L}\
  \bibnamefont {Hughes}}, \ and\ \bibinfo {author} {\bibfnamefont {Shou-Cheng}\
  \bibnamefont {Zhang}},\ }\bibfield  {title} {\enquote {\bibinfo {title}
  {Quantum spin hall effect and topological phase transition in hgte quantum
  wells},}\ }\href@noop {} {\bibfield  {journal} {\bibinfo  {journal}
  {science}\ }\textbf {\bibinfo {volume} {314}},\ \bibinfo {pages} {1757--1761}
  (\bibinfo {year} {2006})}\BibitemShut {NoStop}%
\bibitem [{\citenamefont {Fang}\ \emph {et~al.}(2012)\citenamefont {Fang},
  \citenamefont {Gilbert},\ and\ \citenamefont {Bernevig}}]{Fang_2012}%
  \BibitemOpen
  \bibfield  {author} {\bibinfo {author} {\bibfnamefont {Chen}\ \bibnamefont
  {Fang}}, \bibinfo {author} {\bibfnamefont {Matthew~J.}\ \bibnamefont
  {Gilbert}}, \ and\ \bibinfo {author} {\bibfnamefont {B.~Andrei}\ \bibnamefont
  {Bernevig}},\ }\bibfield  {title} {\enquote {\bibinfo {title} {Bulk
  topological invariants in noninteracting point group symmetric insulators},}\
  }\href {\doibase 10.1103/PhysRevB.86.115112} {\bibfield  {journal} {\bibinfo
  {journal} {Phys. Rev. B}\ }\textbf {\bibinfo {volume} {86}},\ \bibinfo
  {pages} {115112} (\bibinfo {year} {2012})}\BibitemShut {NoStop}%
\bibitem [{\citenamefont {Kane}\ and\ \citenamefont
  {Mele}(2005)}]{KaneMele_Z2}%
  \BibitemOpen
  \bibfield  {author} {\bibinfo {author} {\bibfnamefont {C.~L.}\ \bibnamefont
  {Kane}}\ and\ \bibinfo {author} {\bibfnamefont {E.~J.}\ \bibnamefont
  {Mele}},\ }\bibfield  {title} {\enquote {\bibinfo {title} {${Z}_{2}$
  topological order and the quantum spin hall effect},}\ }\href {\doibase
  10.1103/PhysRevLett.95.146802} {\bibfield  {journal} {\bibinfo  {journal}
  {Phys. Rev. Lett.}\ }\textbf {\bibinfo {volume} {95}},\ \bibinfo {pages}
  {146802} (\bibinfo {year} {2005})}\BibitemShut {NoStop}%
\bibitem [{\citenamefont {Slager}\ \emph {et~al.}(2013)\citenamefont {Slager},
  \citenamefont {Mesaros}, \citenamefont {Juri{\v c}i{\'c}},\ and\
  \citenamefont {Zaanen}}]{Slager13_NatPhys}%
  \BibitemOpen
  \bibfield  {author} {\bibinfo {author} {\bibfnamefont {Robert-Jan}\
  \bibnamefont {Slager}}, \bibinfo {author} {\bibfnamefont {Andrej}\
  \bibnamefont {Mesaros}}, \bibinfo {author} {\bibfnamefont {Vladimir}\
  \bibnamefont {Juri{\v c}i{\'c}}}, \ and\ \bibinfo {author} {\bibfnamefont
  {Jan}\ \bibnamefont {Zaanen}},\ }\bibfield  {title} {\enquote {\bibinfo
  {title} {The space group classification of topological band-insulators},}\
  }\href {http://dx.doi.org/10.1038/nphys2513} {\bibfield  {journal} {\bibinfo
  {journal} {Nat. Phys.}\ }\textbf {\bibinfo {volume} {9}},\ \bibinfo {pages}
  {98} (\bibinfo {year} {2013})}\BibitemShut {NoStop}%
\bibitem [{\citenamefont {Kruthoff}\ \emph {et~al.}(2017)\citenamefont
  {Kruthoff}, \citenamefont {de~Boer}, \citenamefont {van Wezel}, \citenamefont
  {Kane},\ and\ \citenamefont {Slager}}]{Kruthoff_2017}%
  \BibitemOpen
  \bibfield  {author} {\bibinfo {author} {\bibfnamefont {Jorrit}\ \bibnamefont
  {Kruthoff}}, \bibinfo {author} {\bibfnamefont {Jan}\ \bibnamefont {de~Boer}},
  \bibinfo {author} {\bibfnamefont {Jasper}\ \bibnamefont {van Wezel}},
  \bibinfo {author} {\bibfnamefont {Charles~L.}\ \bibnamefont {Kane}}, \ and\
  \bibinfo {author} {\bibfnamefont {Robert-Jan}\ \bibnamefont {Slager}},\
  }\bibfield  {title} {\enquote {\bibinfo {title} {Topological classification
  of crystalline insulators through band structure combinatorics},}\ }\href
  {\doibase 10.1103/PhysRevX.7.041069} {\bibfield  {journal} {\bibinfo
  {journal} {Phys. Rev. X}\ }\textbf {\bibinfo {volume} {7}},\ \bibinfo {pages}
  {041069} (\bibinfo {year} {2017})}\BibitemShut {NoStop}%
\bibitem [{\citenamefont {Po}\ \emph {et~al.}(2017)\citenamefont {Po},
  \citenamefont {Vishwanath},\ and\ \citenamefont
  {Watanabe}}]{PoVishwanath17_NatComm}%
  \BibitemOpen
  \bibfield  {author} {\bibinfo {author} {\bibfnamefont {Hoi~Chun}\
  \bibnamefont {Po}}, \bibinfo {author} {\bibfnamefont {Ashvin}\ \bibnamefont
  {Vishwanath}}, \ and\ \bibinfo {author} {\bibfnamefont {Haruki}\ \bibnamefont
  {Watanabe}},\ }\bibfield  {title} {\enquote {\bibinfo {title} {Symmetry-based
  indicators of band topology in the 230 space groups},}\ }\href {\doibase
  10.1038/s41467-017-00133-2} {\bibfield  {journal} {\bibinfo  {journal} {Nat.
  Commun.}\ }\textbf {\bibinfo {volume} {8}},\ \bibinfo {pages} {50} (\bibinfo
  {year} {2017})}\BibitemShut {NoStop}%
\bibitem [{\citenamefont {Bradlyn}\ \emph {et~al.}(2017)\citenamefont
  {Bradlyn}, \citenamefont {Elcoro}, \citenamefont {Cano}, \citenamefont
  {Vergniory}, \citenamefont {Wang}, \citenamefont {Felser}, \citenamefont
  {Aroyo},\ and\ \citenamefont {Bernevig}}]{Bradlyn17_Nat}%
  \BibitemOpen
  \bibfield  {author} {\bibinfo {author} {\bibfnamefont {Barry}\ \bibnamefont
  {Bradlyn}}, \bibinfo {author} {\bibfnamefont {L.}~\bibnamefont {Elcoro}},
  \bibinfo {author} {\bibfnamefont {Jennifer}\ \bibnamefont {Cano}}, \bibinfo
  {author} {\bibfnamefont {M.~G.}\ \bibnamefont {Vergniory}}, \bibinfo {author}
  {\bibfnamefont {Zhijun}\ \bibnamefont {Wang}}, \bibinfo {author}
  {\bibfnamefont {C.}~\bibnamefont {Felser}}, \bibinfo {author} {\bibfnamefont
  {M.~I.}\ \bibnamefont {Aroyo}}, \ and\ \bibinfo {author} {\bibfnamefont
  {B.~Andrei}\ \bibnamefont {Bernevig}},\ }\bibfield  {title} {\enquote
  {\bibinfo {title} {Topological quantum chemistry},}\ }\href
  {http://dx.doi.org/10.1038/nature23268} {\bibfield  {journal} {\bibinfo
  {journal} {Nature}\ }\textbf {\bibinfo {volume} {547}},\ \bibinfo {pages}
  {298} (\bibinfo {year} {2017})}\BibitemShut {NoStop}%
\bibitem [{\citenamefont {Jotzu}\ \emph {et~al.}(2014)\citenamefont {Jotzu},
  \citenamefont {Messer}, \citenamefont {Desbuquois}, \citenamefont {Lebrat},
  \citenamefont {Uehlinger}, \citenamefont {Greif},\ and\ \citenamefont
  {Esslinger}}]{Jotzu14_Nat}%
  \BibitemOpen
  \bibfield  {author} {\bibinfo {author} {\bibfnamefont {Gregor}\ \bibnamefont
  {Jotzu}}, \bibinfo {author} {\bibfnamefont {Michael}\ \bibnamefont {Messer}},
  \bibinfo {author} {\bibfnamefont {R\'{e}mi}\ \bibnamefont {Desbuquois}},
  \bibinfo {author} {\bibfnamefont {Martin}\ \bibnamefont {Lebrat}}, \bibinfo
  {author} {\bibfnamefont {Thomas}\ \bibnamefont {Uehlinger}}, \bibinfo
  {author} {\bibfnamefont {Daniel}\ \bibnamefont {Greif}}, \ and\ \bibinfo
  {author} {\bibfnamefont {Tilman}\ \bibnamefont {Esslinger}},\ }\bibfield
  {title} {\enquote {\bibinfo {title} {Experimental realization of the
  topological {H}aldane model with ultracold fermions},}\ }\href {\doibase
  10.1038/nature13915} {\bibfield  {journal} {\bibinfo  {journal} {Nature}\
  }\textbf {\bibinfo {volume} {515}},\ \bibinfo {pages} {237} (\bibinfo {year}
  {2014})}\BibitemShut {NoStop}%
\bibitem [{\citenamefont {Aidelsburger}\ \emph {et~al.}(2015)\citenamefont
  {Aidelsburger}, \citenamefont {Lohse}, \citenamefont {Schweizer},
  \citenamefont {Atala}, \citenamefont {Barreiro}, \citenamefont {Nascimbene},
  \citenamefont {Cooper}, \citenamefont {Bloch},\ and\ \citenamefont
  {Goldman}}]{Aidelsburger15_NatPhys}%
  \BibitemOpen
  \bibfield  {author} {\bibinfo {author} {\bibfnamefont {Monika}\ \bibnamefont
  {Aidelsburger}}, \bibinfo {author} {\bibfnamefont {Michael}\ \bibnamefont
  {Lohse}}, \bibinfo {author} {\bibfnamefont {C}~\bibnamefont {Schweizer}},
  \bibinfo {author} {\bibfnamefont {Marcos}\ \bibnamefont {Atala}}, \bibinfo
  {author} {\bibfnamefont {Julio~T}\ \bibnamefont {Barreiro}}, \bibinfo
  {author} {\bibfnamefont {S}~\bibnamefont {Nascimbene}}, \bibinfo {author}
  {\bibfnamefont {NR}~\bibnamefont {Cooper}}, \bibinfo {author} {\bibfnamefont
  {Immanuel}\ \bibnamefont {Bloch}}, \ and\ \bibinfo {author} {\bibfnamefont
  {N}~\bibnamefont {Goldman}},\ }\bibfield  {title} {\enquote {\bibinfo {title}
  {Measuring the {C}hern number of {H}ofstadter bands with ultracold bosonic
  atoms},}\ }\href {https://www.nature.com/articles/nphys3171} {\bibfield
  {journal} {\bibinfo  {journal} {Nat. Phys.}\ }\textbf {\bibinfo {volume}
  {11}},\ \bibinfo {pages} {162--166} (\bibinfo {year} {2015})}\BibitemShut
  {NoStop}%
\bibitem [{\citenamefont {Tran}\ \emph {et~al.}(2017)\citenamefont {Tran},
  \citenamefont {Dauphin}, \citenamefont {Grushin}, \citenamefont {Zoller},\
  and\ \citenamefont {Goldman}}]{Tran17_SciAdv_dichrosim}%
  \BibitemOpen
  \bibfield  {author} {\bibinfo {author} {\bibfnamefont {Duc~Thanh}\
  \bibnamefont {Tran}}, \bibinfo {author} {\bibfnamefont {Alexandre}\
  \bibnamefont {Dauphin}}, \bibinfo {author} {\bibfnamefont {Adolfo~G}\
  \bibnamefont {Grushin}}, \bibinfo {author} {\bibfnamefont {Peter}\
  \bibnamefont {Zoller}}, \ and\ \bibinfo {author} {\bibfnamefont {Nathan}\
  \bibnamefont {Goldman}},\ }\bibfield  {title} {\enquote {\bibinfo {title}
  {Probing topology by “heating”: Quantized circular dichroism in ultracold
  atoms},}\ }\href@noop {} {\bibfield  {journal} {\bibinfo  {journal} {Science
  advances}\ }\textbf {\bibinfo {volume} {3}},\ \bibinfo {pages} {e1701207}
  (\bibinfo {year} {2017})}\BibitemShut {NoStop}%
\bibitem [{\citenamefont {Asteria}\ \emph {et~al.}(2019)\citenamefont
  {Asteria}, \citenamefont {Tran}, \citenamefont {Ozawa}, \citenamefont
  {Tarnowski}, \citenamefont {Rem}, \citenamefont {Flashner}, \citenamefont
  {Sengstock}, \citenamefont {Goldman},\ and\ \citenamefont
  {Weitenberg}}]{Asteria19_NatPhys}%
  \BibitemOpen
  \bibfield  {author} {\bibinfo {author} {\bibfnamefont {L.}~\bibnamefont
  {Asteria}}, \bibinfo {author} {\bibfnamefont {D.T.}\ \bibnamefont {Tran}},
  \bibinfo {author} {\bibfnamefont {T.}~\bibnamefont {Ozawa}}, \bibinfo
  {author} {\bibfnamefont {M.}~\bibnamefont {Tarnowski}}, \bibinfo {author}
  {\bibfnamefont {B.~S.}\ \bibnamefont {Rem}}, \bibinfo {author} {\bibfnamefont
  {N.}~\bibnamefont {Flashner}}, \bibinfo {author} {\bibfnamefont
  {K.}~\bibnamefont {Sengstock}}, \bibinfo {author} {\bibfnamefont
  {N.}~\bibnamefont {Goldman}}, \ and\ \bibinfo {author} {\bibfnamefont
  {C.}~\bibnamefont {Weitenberg}},\ }\bibfield  {title} {\enquote {\bibinfo
  {title} {Measuring quantized circular dichroism in ultracold topological
  matter},}\ }\href {\doibase 10.1038/s41567-019-0417-8} {\bibfield  {journal}
  {\bibinfo  {journal} {Nat. Phys}\ }\textbf {\bibinfo {volume} {15}},\
  \bibinfo {pages} {449} (\bibinfo {year} {2019})}\BibitemShut {NoStop}%
\bibitem [{\citenamefont {Kemp}\ \emph {et~al.}(2022)\citenamefont {Kemp},
  \citenamefont {Cooper},\ and\ \citenamefont {\"Unal}}]{Kemp22_PRR}%
  \BibitemOpen
  \bibfield  {author} {\bibinfo {author} {\bibfnamefont {Cameron J.~D.}\
  \bibnamefont {Kemp}}, \bibinfo {author} {\bibfnamefont {Nigel~R.}\
  \bibnamefont {Cooper}}, \ and\ \bibinfo {author} {\bibfnamefont {F.~Nur}\
  \bibnamefont {\"Unal}},\ }\bibfield  {title} {\enquote {\bibinfo {title}
  {Nested-sphere description of the $n$-level chern number and the generalized
  bloch hypersphere},}\ }\href {\doibase 10.1103/PhysRevResearch.4.023120}
  {\bibfield  {journal} {\bibinfo  {journal} {Phys. Rev. Res.}\ }\textbf
  {\bibinfo {volume} {4}},\ \bibinfo {pages} {023120} (\bibinfo {year}
  {2022})}\BibitemShut {NoStop}%
\bibitem [{\citenamefont {Tan}\ \emph {et~al.}(2019)\citenamefont {Tan},
  \citenamefont {Zhang}, \citenamefont {Yang}, \citenamefont {Chu},
  \citenamefont {Zhu}, \citenamefont {Li}, \citenamefont {Yang}, \citenamefont
  {Song}, \citenamefont {Han}, \citenamefont {Li}, \citenamefont {Dong},
  \citenamefont {Yu}, \citenamefont {Yan}, \citenamefont {Zhu},\ and\
  \citenamefont {Yu}}]{TanYu19_PRL_QGTmeasurement}%
  \BibitemOpen
  \bibfield  {author} {\bibinfo {author} {\bibfnamefont {Xinsheng}\
  \bibnamefont {Tan}}, \bibinfo {author} {\bibfnamefont {Dan-Wei}\ \bibnamefont
  {Zhang}}, \bibinfo {author} {\bibfnamefont {Zhen}\ \bibnamefont {Yang}},
  \bibinfo {author} {\bibfnamefont {Ji}~\bibnamefont {Chu}}, \bibinfo {author}
  {\bibfnamefont {Yan-Qing}\ \bibnamefont {Zhu}}, \bibinfo {author}
  {\bibfnamefont {Danyu}\ \bibnamefont {Li}}, \bibinfo {author} {\bibfnamefont
  {Xiaopei}\ \bibnamefont {Yang}}, \bibinfo {author} {\bibfnamefont {Shuqing}\
  \bibnamefont {Song}}, \bibinfo {author} {\bibfnamefont {Zhikun}\ \bibnamefont
  {Han}}, \bibinfo {author} {\bibfnamefont {Zhiyuan}\ \bibnamefont {Li}},
  \bibinfo {author} {\bibfnamefont {Yuqian}\ \bibnamefont {Dong}}, \bibinfo
  {author} {\bibfnamefont {Hai-Feng}\ \bibnamefont {Yu}}, \bibinfo {author}
  {\bibfnamefont {Hui}\ \bibnamefont {Yan}}, \bibinfo {author} {\bibfnamefont
  {Shi-Liang}\ \bibnamefont {Zhu}}, \ and\ \bibinfo {author} {\bibfnamefont
  {Yang}\ \bibnamefont {Yu}},\ }\bibfield  {title} {\enquote {\bibinfo {title}
  {Experimental measurement of the quantum metric tensor and related
  topological phase transition with a superconducting qubit},}\ }\href
  {\doibase 10.1103/PhysRevLett.122.210401} {\bibfield  {journal} {\bibinfo
  {journal} {Phys. Rev. Lett.}\ }\textbf {\bibinfo {volume} {122}},\ \bibinfo
  {pages} {210401} (\bibinfo {year} {2019})}\BibitemShut {NoStop}%
\bibitem [{\citenamefont {Xu}\ \emph {et~al.}(2022)\citenamefont {Xu},
  \citenamefont {Zheng},\ and\ \citenamefont {Zhai}}]{Zhai2022}%
  \BibitemOpen
  \bibfield  {author} {\bibinfo {author} {\bibfnamefont {Peng}\ \bibnamefont
  {Xu}}, \bibinfo {author} {\bibfnamefont {Wei}\ \bibnamefont {Zheng}}, \ and\
  \bibinfo {author} {\bibfnamefont {Hui}\ \bibnamefont {Zhai}},\ }\bibfield
  {title} {\enquote {\bibinfo {title} {Topological micromotion of floquet
  quantum systems},}\ }\href {\doibase 10.1103/PhysRevB.105.045139} {\bibfield
  {journal} {\bibinfo  {journal} {Phys. Rev. B}\ }\textbf {\bibinfo {volume}
  {105}},\ \bibinfo {pages} {045139} (\bibinfo {year} {2022})}\BibitemShut
  {NoStop}%
\bibitem [{\citenamefont {Jangjan}\ \emph {et~al.}(2022)\citenamefont
  {Jangjan}, \citenamefont {Foa~Torres},\ and\ \citenamefont
  {Hosseini}}]{Vahid2022}%
  \BibitemOpen
  \bibfield  {author} {\bibinfo {author} {\bibfnamefont {Milad}\ \bibnamefont
  {Jangjan}}, \bibinfo {author} {\bibfnamefont {Luis E.~F.}\ \bibnamefont
  {Foa~Torres}}, \ and\ \bibinfo {author} {\bibfnamefont {Mir~Vahid}\
  \bibnamefont {Hosseini}},\ }\bibfield  {title} {\enquote {\bibinfo {title}
  {Floquet topological phase transitions in a periodically quenched dimer},}\
  }\href {\doibase 10.1103/PhysRevB.106.224306} {\bibfield  {journal} {\bibinfo
   {journal} {Phys. Rev. B}\ }\textbf {\bibinfo {volume} {106}},\ \bibinfo
  {pages} {224306} (\bibinfo {year} {2022})}\BibitemShut {NoStop}%
\bibitem [{\citenamefont {Jangjan}\ and\ \citenamefont
  {Hosseini}(2020)}]{jangjan2020}%
  \BibitemOpen
  \bibfield  {author} {\bibinfo {author} {\bibfnamefont {Milad}\ \bibnamefont
  {Jangjan}}\ and\ \bibinfo {author} {\bibfnamefont {Mir~Vahid}\ \bibnamefont
  {Hosseini}},\ }\bibfield  {title} {\enquote {\bibinfo {title} {Floquet
  engineering of topological metal states and hybridization of edge states with
  bulk states in dimerized two-leg ladders},}\ }\href {\doibase
  10.1038/s41598-020-71196-3} {\bibfield  {journal} {\bibinfo  {journal}
  {Scientific Reports}\ }\textbf {\bibinfo {volume} {10}},\ \bibinfo {pages}
  {14256} (\bibinfo {year} {2020})}\BibitemShut {NoStop}%
\bibitem [{\citenamefont {Roy}\ and\ \citenamefont {Harper}(2017)}]{Roy_PRB}%
  \BibitemOpen
  \bibfield  {author} {\bibinfo {author} {\bibfnamefont {Rahul}\ \bibnamefont
  {Roy}}\ and\ \bibinfo {author} {\bibfnamefont {Fenner}\ \bibnamefont
  {Harper}},\ }\bibfield  {title} {\enquote {\bibinfo {title} {Periodic table
  for floquet topological insulators},}\ }\href {\doibase
  10.1103/PhysRevB.96.155118} {\bibfield  {journal} {\bibinfo  {journal} {Phys.
  Rev. B}\ }\textbf {\bibinfo {volume} {96}},\ \bibinfo {pages} {155118}
  (\bibinfo {year} {2017})}\BibitemShut {NoStop}%
\bibitem [{\citenamefont {Oka}\ and\ \citenamefont {Aoki}(2009)}]{Oka_PRB}%
  \BibitemOpen
  \bibfield  {author} {\bibinfo {author} {\bibfnamefont {Takashi}\ \bibnamefont
  {Oka}}\ and\ \bibinfo {author} {\bibfnamefont {Hideo}\ \bibnamefont {Aoki}},\
  }\bibfield  {title} {\enquote {\bibinfo {title} {Photovoltaic hall effect in
  graphene},}\ }\href {\doibase 10.1103/PhysRevB.79.081406} {\bibfield
  {journal} {\bibinfo  {journal} {Phys. Rev. B}\ }\textbf {\bibinfo {volume}
  {79}},\ \bibinfo {pages} {081406} (\bibinfo {year} {2009})}\BibitemShut
  {NoStop}%
\bibitem [{\citenamefont {Kitagawa}\ \emph {et~al.}(2010)\citenamefont
  {Kitagawa}, \citenamefont {Berg}, \citenamefont {Rudner},\ and\ \citenamefont
  {Demler}}]{Kit_PRB}%
  \BibitemOpen
  \bibfield  {author} {\bibinfo {author} {\bibfnamefont {Takuya}\ \bibnamefont
  {Kitagawa}}, \bibinfo {author} {\bibfnamefont {Erez}\ \bibnamefont {Berg}},
  \bibinfo {author} {\bibfnamefont {Mark}\ \bibnamefont {Rudner}}, \ and\
  \bibinfo {author} {\bibfnamefont {Eugene}\ \bibnamefont {Demler}},\
  }\bibfield  {title} {\enquote {\bibinfo {title} {Topological characterization
  of periodically driven quantum systems},}\ }\href {\doibase
  10.1103/PhysRevB.82.235114} {\bibfield  {journal} {\bibinfo  {journal} {Phys.
  Rev. B}\ }\textbf {\bibinfo {volume} {82}},\ \bibinfo {pages} {235114}
  (\bibinfo {year} {2010})}\BibitemShut {NoStop}%
\bibitem [{\citenamefont {Rudner}\ \emph {et~al.}(2013)\citenamefont {Rudner},
  \citenamefont {Lindner}, \citenamefont {Berg},\ and\ \citenamefont
  {Levin}}]{Rudner_PRX}%
  \BibitemOpen
  \bibfield  {author} {\bibinfo {author} {\bibfnamefont {Mark~S.}\ \bibnamefont
  {Rudner}}, \bibinfo {author} {\bibfnamefont {Netanel~H.}\ \bibnamefont
  {Lindner}}, \bibinfo {author} {\bibfnamefont {Erez}\ \bibnamefont {Berg}}, \
  and\ \bibinfo {author} {\bibfnamefont {Michael}\ \bibnamefont {Levin}},\
  }\bibfield  {title} {\enquote {\bibinfo {title} {Anomalous edge states and
  the bulk-edge correspondence for periodically driven two-dimensional
  systems},}\ }\href {\doibase 10.1103/PhysRevX.3.031005} {\bibfield  {journal}
  {\bibinfo  {journal} {Phys. Rev. X}\ }\textbf {\bibinfo {volume} {3}},\
  \bibinfo {pages} {031005} (\bibinfo {year} {2013})}\BibitemShut {NoStop}%
\bibitem [{\citenamefont {\"Unal}\ \emph
  {et~al.}(2019{\natexlab{a}})\citenamefont {\"Unal}, \citenamefont
  {Seradjeh},\ and\ \citenamefont {Eckardt}}]{Nur_PRL_HowToMeasure}%
  \BibitemOpen
  \bibfield  {author} {\bibinfo {author} {\bibfnamefont {F.~Nur}\ \bibnamefont
  {\"Unal}}, \bibinfo {author} {\bibfnamefont {Babak}\ \bibnamefont
  {Seradjeh}}, \ and\ \bibinfo {author} {\bibfnamefont {Andr\'e}\ \bibnamefont
  {Eckardt}},\ }\bibfield  {title} {\enquote {\bibinfo {title} {How to directly
  measure floquet topological invariants in optical lattices},}\ }\href
  {\doibase 10.1103/PhysRevLett.122.253601} {\bibfield  {journal} {\bibinfo
  {journal} {Phys. Rev. Lett.}\ }\textbf {\bibinfo {volume} {122}},\ \bibinfo
  {pages} {253601} (\bibinfo {year} {2019}{\natexlab{a}})}\BibitemShut
  {NoStop}%
\bibitem [{\citenamefont {Wintersperger}\ \emph {et~al.}(2020)\citenamefont
  {Wintersperger}, \citenamefont {Braun}, \citenamefont {{\"U}nal},
  \citenamefont {Eckardt}, \citenamefont {Liberto}, \citenamefont {Goldman},
  \citenamefont {Bloch},\ and\ \citenamefont
  {Aidelsburger}}]{wintersperger2020realization}%
  \BibitemOpen
  \bibfield  {author} {\bibinfo {author} {\bibfnamefont {Karen}\ \bibnamefont
  {Wintersperger}}, \bibinfo {author} {\bibfnamefont {Christoph}\ \bibnamefont
  {Braun}}, \bibinfo {author} {\bibfnamefont {F~Nur}\ \bibnamefont {{\"U}nal}},
  \bibinfo {author} {\bibfnamefont {Andr{\'e}}\ \bibnamefont {Eckardt}},
  \bibinfo {author} {\bibfnamefont {Marco~Di}\ \bibnamefont {Liberto}},
  \bibinfo {author} {\bibfnamefont {Nathan}\ \bibnamefont {Goldman}}, \bibinfo
  {author} {\bibfnamefont {Immanuel}\ \bibnamefont {Bloch}}, \ and\ \bibinfo
  {author} {\bibfnamefont {Monika}\ \bibnamefont {Aidelsburger}},\ }\bibfield
  {title} {\enquote {\bibinfo {title} {Realization of an anomalous floquet
  topological system with ultracold atoms},}\ }\href
  {https://www.nature.com/articles/s41567-020-0949-y#citeas} {\bibfield
  {journal} {\bibinfo  {journal} {Nature Physics}\ }\textbf {\bibinfo {volume}
  {16}},\ \bibinfo {pages} {1058--1063} (\bibinfo {year} {2020})}\BibitemShut
  {NoStop}%
\bibitem [{\citenamefont {Wang}\ \emph {et~al.}(2017)\citenamefont {Wang},
  \citenamefont {Zhang}, \citenamefont {Chen}, \citenamefont {Yu},\ and\
  \citenamefont {Zhai}}]{Wang_2017}%
  \BibitemOpen
  \bibfield  {author} {\bibinfo {author} {\bibfnamefont {Ce}~\bibnamefont
  {Wang}}, \bibinfo {author} {\bibfnamefont {Pengfei}\ \bibnamefont {Zhang}},
  \bibinfo {author} {\bibfnamefont {Xin}\ \bibnamefont {Chen}}, \bibinfo
  {author} {\bibfnamefont {Jinlong}\ \bibnamefont {Yu}}, \ and\ \bibinfo
  {author} {\bibfnamefont {Hui}\ \bibnamefont {Zhai}},\ }\bibfield  {title}
  {\enquote {\bibinfo {title} {Scheme to measure the topological number of a
  chern insulator from quench dynamics},}\ }\href {\doibase
  10.1103/PhysRevLett.118.185701} {\bibfield  {journal} {\bibinfo  {journal}
  {Phys. Rev. Lett.}\ }\textbf {\bibinfo {volume} {118}},\ \bibinfo {pages}
  {185701} (\bibinfo {year} {2017})}\BibitemShut {NoStop}%
\bibitem [{\citenamefont {Tarnowski}\ \emph {et~al.}(2019)\citenamefont
  {Tarnowski}, \citenamefont {{\"U}nal}, \citenamefont {Fl{\"a}schner},
  \citenamefont {Rem}, \citenamefont {Eckardt}, \citenamefont {Sengstock},\
  and\ \citenamefont {Weitenberg}}]{Tarnowski19_NatCom}%
  \BibitemOpen
  \bibfield  {author} {\bibinfo {author} {\bibfnamefont {Matthias}\
  \bibnamefont {Tarnowski}}, \bibinfo {author} {\bibfnamefont {F~Nur}\
  \bibnamefont {{\"U}nal}}, \bibinfo {author} {\bibfnamefont {Nick}\
  \bibnamefont {Fl{\"a}schner}}, \bibinfo {author} {\bibfnamefont {Benno~S}\
  \bibnamefont {Rem}}, \bibinfo {author} {\bibfnamefont {Andr{\'e}}\
  \bibnamefont {Eckardt}}, \bibinfo {author} {\bibfnamefont {Klaus}\
  \bibnamefont {Sengstock}}, \ and\ \bibinfo {author} {\bibfnamefont
  {Christof}\ \bibnamefont {Weitenberg}},\ }\bibfield  {title} {\enquote
  {\bibinfo {title} {Measuring topology from dynamics by obtaining the {C}hern
  number from a linking number},}\ }\href
  {https://www.nature.com/articles/s41467-019-09668-y} {\bibfield  {journal}
  {\bibinfo  {journal} {Nat. Commun.}\ }\textbf {\bibinfo {volume} {10}}
  (\bibinfo {year} {2019})}\BibitemShut {NoStop}%
\bibitem [{\citenamefont {\"Unal}\ \emph
  {et~al.}(2019{\natexlab{b}})\citenamefont {\"Unal}, \citenamefont {Eckardt},\
  and\ \citenamefont {Slager}}]{Unal19_PRR_hopf}%
  \BibitemOpen
  \bibfield  {author} {\bibinfo {author} {\bibfnamefont {F.~Nur}\ \bibnamefont
  {\"Unal}}, \bibinfo {author} {\bibfnamefont {Andr\'e}\ \bibnamefont
  {Eckardt}}, \ and\ \bibinfo {author} {\bibfnamefont {Robert-Jan}\
  \bibnamefont {Slager}},\ }\bibfield  {title} {\enquote {\bibinfo {title}
  {Hopf characterization of two-dimensional floquet topological insulators},}\
  }\href {\doibase 10.1103/PhysRevResearch.1.022003} {\bibfield  {journal}
  {\bibinfo  {journal} {Phys. Rev. Research}\ }\textbf {\bibinfo {volume}
  {1}},\ \bibinfo {pages} {022003} (\bibinfo {year}
  {2019}{\natexlab{b}})}\BibitemShut {NoStop}%
\bibitem [{\citenamefont {Hu}\ and\ \citenamefont
  {Zhao}(2020)}]{HuZhao20_PRL_hopf}%
  \BibitemOpen
  \bibfield  {author} {\bibinfo {author} {\bibfnamefont {Haiping}\ \bibnamefont
  {Hu}}\ and\ \bibinfo {author} {\bibfnamefont {Erhai}\ \bibnamefont {Zhao}},\
  }\bibfield  {title} {\enquote {\bibinfo {title} {Topological invariants for
  quantum quench dynamics from unitary evolution},}\ }\href {\doibase
  10.1103/PhysRevLett.124.160402} {\bibfield  {journal} {\bibinfo  {journal}
  {Phys. Rev. Lett.}\ }\textbf {\bibinfo {volume} {124}},\ \bibinfo {pages}
  {160402} (\bibinfo {year} {2020})}\BibitemShut {NoStop}%
\bibitem [{\citenamefont {\"Unal}\ \emph {et~al.}(2020)\citenamefont {\"Unal},
  \citenamefont {Bouhon},\ and\ \citenamefont {Slager}}]{Unal_2020}%
  \BibitemOpen
  \bibfield  {author} {\bibinfo {author} {\bibfnamefont {F.~Nur}\ \bibnamefont
  {\"Unal}}, \bibinfo {author} {\bibfnamefont {Adrien}\ \bibnamefont {Bouhon}},
  \ and\ \bibinfo {author} {\bibfnamefont {Robert-Jan}\ \bibnamefont
  {Slager}},\ }\bibfield  {title} {\enquote {\bibinfo {title} {Topological
  euler class as a dynamical observable in optical lattices},}\ }\href
  {\doibase 10.1103/PhysRevLett.125.053601} {\bibfield  {journal} {\bibinfo
  {journal} {Phys. Rev. Lett.}\ }\textbf {\bibinfo {volume} {125}},\ \bibinfo
  {pages} {053601} (\bibinfo {year} {2020})}\BibitemShut {NoStop}%
\bibitem [{\citenamefont {Slager}\ \emph {et~al.}(2022)\citenamefont {Slager},
  \citenamefont {Bouhon},\ and\ \citenamefont
  {\"Unal}}]{Slager22_arXivAnomEuFloq}%
  \BibitemOpen
  \bibfield  {author} {\bibinfo {author} {\bibfnamefont {Robert-Jan}\
  \bibnamefont {Slager}}, \bibinfo {author} {\bibfnamefont {Adrien}\
  \bibnamefont {Bouhon}}, \ and\ \bibinfo {author} {\bibfnamefont {F.~Nur}\
  \bibnamefont {\"Unal}},\ }\href {\doibase 10.48550/ARXIV.2208.12824}
  {\enquote {\bibinfo {title} {Floquet multi-gap topology: Non-abelian braiding
  and anomalous dirac string phase},}\ } (\bibinfo {year} {2022}),\ \Eprint
  {http://arxiv.org/abs/2208.12824} {arXiv:2208.12824} \BibitemShut {NoStop}%
\bibitem [{\citenamefont {Ahn}\ \emph {et~al.}(2019)\citenamefont {Ahn},
  \citenamefont {Park},\ and\ \citenamefont {Yang}}]{Ahn2019}%
  \BibitemOpen
  \bibfield  {author} {\bibinfo {author} {\bibfnamefont {Junyeong}\
  \bibnamefont {Ahn}}, \bibinfo {author} {\bibfnamefont {Sungjoon}\
  \bibnamefont {Park}}, \ and\ \bibinfo {author} {\bibfnamefont {Bohm-Jung}\
  \bibnamefont {Yang}},\ }\bibfield  {title} {\enquote {\bibinfo {title}
  {Failure of nielsen-ninomiya theorem and fragile topology in two-dimensional
  systems with space-time inversion symmetry: Application to twisted bilayer
  graphene at magic angle},}\ }\href {\doibase 10.1103/PhysRevX.9.021013}
  {\bibfield  {journal} {\bibinfo  {journal} {Phys. Rev. X}\ }\textbf {\bibinfo
  {volume} {9}},\ \bibinfo {pages} {021013} (\bibinfo {year}
  {2019})}\BibitemShut {NoStop}%
\bibitem [{\citenamefont {Bouhon}\ \emph
  {et~al.}(2020{\natexlab{a}})\citenamefont {Bouhon}, \citenamefont {Wu},
  \citenamefont {Slager}, \citenamefont {Weng}, \citenamefont {Yazyev},\ and\
  \citenamefont {Bzdu{\v s}ek}}]{bouhon2019nonabelian}%
  \BibitemOpen
  \bibfield  {author} {\bibinfo {author} {\bibfnamefont {Adrien}\ \bibnamefont
  {Bouhon}}, \bibinfo {author} {\bibfnamefont {QuanSheng}\ \bibnamefont {Wu}},
  \bibinfo {author} {\bibfnamefont {Robert-Jan}\ \bibnamefont {Slager}},
  \bibinfo {author} {\bibfnamefont {Hongming}\ \bibnamefont {Weng}}, \bibinfo
  {author} {\bibfnamefont {Oleg~V.}\ \bibnamefont {Yazyev}}, \ and\ \bibinfo
  {author} {\bibfnamefont {Tom{\'a}{\v s}}\ \bibnamefont {Bzdu{\v s}ek}},\
  }\bibfield  {title} {\enquote {\bibinfo {title} {Non-abelian reciprocal
  braiding of weyl points and its manifestation in zrte},}\ }\href {\doibase
  10.1038/s41567-020-0967-9} {\bibfield  {journal} {\bibinfo  {journal} {Nature
  Physics}\ }\textbf {\bibinfo {volume} {16}},\ \bibinfo {pages} {1137--1143}
  (\bibinfo {year} {2020}{\natexlab{a}})}\BibitemShut {NoStop}%
\bibitem [{\citenamefont {Bouhon}\ \emph
  {et~al.}(2020{\natexlab{b}})\citenamefont {Bouhon}, \citenamefont
  {Bzdu\v{s}ek},\ and\ \citenamefont {Slager}}]{bouhonGeometric2020}%
  \BibitemOpen
  \bibfield  {author} {\bibinfo {author} {\bibfnamefont {Adrien}\ \bibnamefont
  {Bouhon}}, \bibinfo {author} {\bibfnamefont {Tom\'a\v{s}}\ \bibnamefont
  {Bzdu\v{s}ek}}, \ and\ \bibinfo {author} {\bibfnamefont {Robert-Jan}\
  \bibnamefont {Slager}},\ }\bibfield  {title} {\enquote {\bibinfo {title}
  {Geometric approach to fragile topology beyond symmetry indicators},}\ }\href
  {\doibase 10.1103/PhysRevB.102.115135} {\bibfield  {journal} {\bibinfo
  {journal} {Phys. Rev. B}\ }\textbf {\bibinfo {volume} {102}},\ \bibinfo
  {pages} {115135} (\bibinfo {year} {2020}{\natexlab{b}})}\BibitemShut
  {NoStop}%
\bibitem [{\citenamefont {Jiang}\ \emph
  {et~al.}(2021{\natexlab{a}})\citenamefont {Jiang}, \citenamefont {Bouhon},
  \citenamefont {Lin}, \citenamefont {Zhou}, \citenamefont {Hou}, \citenamefont
  {Li}, \citenamefont {Slager},\ and\ \citenamefont {Jiang}}]{Jiang2021}%
  \BibitemOpen
  \bibfield  {author} {\bibinfo {author} {\bibfnamefont {Bin}\ \bibnamefont
  {Jiang}}, \bibinfo {author} {\bibfnamefont {Adrien}\ \bibnamefont {Bouhon}},
  \bibinfo {author} {\bibfnamefont {Zhi-Kang}\ \bibnamefont {Lin}}, \bibinfo
  {author} {\bibfnamefont {Xiaoxi}\ \bibnamefont {Zhou}}, \bibinfo {author}
  {\bibfnamefont {Bo}~\bibnamefont {Hou}}, \bibinfo {author} {\bibfnamefont
  {Feng}\ \bibnamefont {Li}}, \bibinfo {author} {\bibfnamefont {Robert-Jan}\
  \bibnamefont {Slager}}, \ and\ \bibinfo {author} {\bibfnamefont {Jian-Hua}\
  \bibnamefont {Jiang}},\ }\bibfield  {title} {\enquote {\bibinfo {title}
  {Experimental observation of non-abelian topological acoustic semimetals and
  their phase transitions},}\ }\href {\doibase 10.1038/s41567-021-01340-x}
  {\bibfield  {journal} {\bibinfo  {journal} {Nature Physics}\ }\textbf
  {\bibinfo {volume} {17}},\ \bibinfo {pages} {1239--1246} (\bibinfo {year}
  {2021}{\natexlab{a}})}\BibitemShut {NoStop}%
\bibitem [{\citenamefont {Jiang}\ \emph
  {et~al.}(2021{\natexlab{b}})\citenamefont {Jiang}, \citenamefont {Guo},
  \citenamefont {Zhang}, \citenamefont {Zhang}, \citenamefont {Yang},\ and\
  \citenamefont {Chan}}]{Jiang1Dexp}%
  \BibitemOpen
  \bibfield  {author} {\bibinfo {author} {\bibfnamefont {Tianshu}\ \bibnamefont
  {Jiang}}, \bibinfo {author} {\bibfnamefont {Qinghua}\ \bibnamefont {Guo}},
  \bibinfo {author} {\bibfnamefont {Ruo-Yang}\ \bibnamefont {Zhang}}, \bibinfo
  {author} {\bibfnamefont {Zhao-Qing}\ \bibnamefont {Zhang}}, \bibinfo {author}
  {\bibfnamefont {Biao}\ \bibnamefont {Yang}}, \ and\ \bibinfo {author}
  {\bibfnamefont {C.~T.}\ \bibnamefont {Chan}},\ }\bibfield  {title} {\enquote
  {\bibinfo {title} {Four-band non-abelian topological insulator and its
  experimental realization},}\ }\href {\doibase 10.1038/s41467-021-26763-1}
  {\bibfield  {journal} {\bibinfo  {journal} {Nature Communications}\ }\textbf
  {\bibinfo {volume} {12}},\ \bibinfo {pages} {6471} (\bibinfo {year}
  {2021}{\natexlab{b}})}\BibitemShut {NoStop}%
\bibitem [{\citenamefont {Goldman}\ and\ \citenamefont
  {Dalibard}(2014)}]{Goldman_PRX}%
  \BibitemOpen
  \bibfield  {author} {\bibinfo {author} {\bibfnamefont {N.}~\bibnamefont
  {Goldman}}\ and\ \bibinfo {author} {\bibfnamefont {J.}~\bibnamefont
  {Dalibard}},\ }\bibfield  {title} {\enquote {\bibinfo {title} {Periodically
  driven quantum systems: Effective hamiltonians and engineered gauge
  fields},}\ }\href {\doibase 10.1103/PhysRevX.4.031027} {\bibfield  {journal}
  {\bibinfo  {journal} {Phys. Rev. X}\ }\textbf {\bibinfo {volume} {4}},\
  \bibinfo {pages} {031027} (\bibinfo {year} {2014})}\BibitemShut {NoStop}%
\bibitem [{\citenamefont {Eckardt}(2017)}]{Eckardt_RevModPhys}%
  \BibitemOpen
  \bibfield  {author} {\bibinfo {author} {\bibfnamefont {Andr\'e}\ \bibnamefont
  {Eckardt}},\ }\bibfield  {title} {\enquote {\bibinfo {title} {Colloquium:
  Atomic quantum gases in periodically driven optical lattices},}\ }\href
  {\doibase 10.1103/RevModPhys.89.011004} {\bibfield  {journal} {\bibinfo
  {journal} {Rev. Mod. Phys.}\ }\textbf {\bibinfo {volume} {89}},\ \bibinfo
  {pages} {011004} (\bibinfo {year} {2017})}\BibitemShut {NoStop}%
\bibitem [{\citenamefont {Bukov}\ \emph {et~al.}(2015)\citenamefont {Bukov},
  \citenamefont {D'Alessio},\ and\ \citenamefont
  {Polkovnikov}}]{Bukov_AdvPhys}%
  \BibitemOpen
  \bibfield  {author} {\bibinfo {author} {\bibfnamefont {Marin}\ \bibnamefont
  {Bukov}}, \bibinfo {author} {\bibfnamefont {Luca}\ \bibnamefont {D'Alessio}},
  \ and\ \bibinfo {author} {\bibfnamefont {Anatoli}\ \bibnamefont
  {Polkovnikov}},\ }\bibfield  {title} {\enquote {\bibinfo {title} {Universal
  high-frequency behavior of periodically driven systems: from dynamical
  stabilization to floquet engineering},}\ }\href {\doibase
  10.1080/00018732.2015.1055918} {\bibfield  {journal} {\bibinfo  {journal}
  {Advances in Physics}\ }\textbf {\bibinfo {volume} {64}},\ \bibinfo {pages}
  {139--226} (\bibinfo {year} {2015})}\BibitemShut {NoStop}%
\bibitem [{\citenamefont {Cooper}\ \emph {et~al.}(2019)\citenamefont {Cooper},
  \citenamefont {Dalibard},\ and\ \citenamefont {Spielman}}]{Cooper19_RMP}%
  \BibitemOpen
  \bibfield  {author} {\bibinfo {author} {\bibfnamefont {N.~R.}\ \bibnamefont
  {Cooper}}, \bibinfo {author} {\bibfnamefont {J.}~\bibnamefont {Dalibard}}, \
  and\ \bibinfo {author} {\bibfnamefont {I.~B.}\ \bibnamefont {Spielman}},\
  }\bibfield  {title} {\enquote {\bibinfo {title} {Topological bands for
  ultracold atoms},}\ }\href {\doibase 10.1103/RevModPhys.91.015005} {\bibfield
   {journal} {\bibinfo  {journal} {Rev. Mod. Phys.}\ }\textbf {\bibinfo
  {volume} {91}},\ \bibinfo {pages} {015005} (\bibinfo {year}
  {2019})}\BibitemShut {NoStop}%
\bibitem [{\citenamefont {Wang}\ \emph {et~al.}(2018)\citenamefont {Wang},
  \citenamefont {\"Unal},\ and\ \citenamefont {Eckardt}}]{WangUnal_18_PRL}%
  \BibitemOpen
  \bibfield  {author} {\bibinfo {author} {\bibfnamefont {Botao}\ \bibnamefont
  {Wang}}, \bibinfo {author} {\bibfnamefont {F.~Nur}\ \bibnamefont {\"Unal}}, \
  and\ \bibinfo {author} {\bibfnamefont {Andr\'e}\ \bibnamefont {Eckardt}},\
  }\bibfield  {title} {\enquote {\bibinfo {title} {Floquet engineering of
  optical solenoids and quantized charge pumping along tailored paths in
  two-dimensional {C}hern insulators},}\ }\href {\doibase
  10.1103/PhysRevLett.120.243602} {\bibfield  {journal} {\bibinfo  {journal}
  {Phys. Rev. Lett.}\ }\textbf {\bibinfo {volume} {120}},\ \bibinfo {pages}
  {243602} (\bibinfo {year} {2018})}\BibitemShut {NoStop}%
\bibitem [{\citenamefont {Ra\ifmmode \check{c}\else
  \v{c}\fi{}i\ifmmode~\bar{u}\else \={u}\fi{}nas}\ \emph
  {et~al.}(2018)\citenamefont {Ra\ifmmode \check{c}\else
  \v{c}\fi{}i\ifmmode~\bar{u}\else \={u}\fi{}nas}, \citenamefont {\"Unal},
  \citenamefont {Anisimovas},\ and\ \citenamefont
  {Eckardt}}]{RaciunasUnal_18_PRA}%
  \BibitemOpen
  \bibfield  {author} {\bibinfo {author} {\bibfnamefont {Mantas}\ \bibnamefont
  {Ra\ifmmode \check{c}\else \v{c}\fi{}i\ifmmode~\bar{u}\else \={u}\fi{}nas}},
  \bibinfo {author} {\bibfnamefont {F.~Nur}\ \bibnamefont {\"Unal}}, \bibinfo
  {author} {\bibfnamefont {Egidijus}\ \bibnamefont {Anisimovas}}, \ and\
  \bibinfo {author} {\bibfnamefont {Andr\'e}\ \bibnamefont {Eckardt}},\
  }\bibfield  {title} {\enquote {\bibinfo {title} {Creating, probing, and
  manipulating fractionally charged excitations of fractional chern insulators
  in optical lattices},}\ }\href {\doibase 10.1103/PhysRevA.98.063621}
  {\bibfield  {journal} {\bibinfo  {journal} {Phys. Rev. A}\ }\textbf {\bibinfo
  {volume} {98}},\ \bibinfo {pages} {063621} (\bibinfo {year}
  {2018})}\BibitemShut {NoStop}%
\bibitem [{\citenamefont {Reichl}\ and\ \citenamefont
  {Mueller}(2014)}]{Reichl14_PRA}%
  \BibitemOpen
  \bibfield  {author} {\bibinfo {author} {\bibfnamefont {Matthew~D.}\
  \bibnamefont {Reichl}}\ and\ \bibinfo {author} {\bibfnamefont {Erich~J.}\
  \bibnamefont {Mueller}},\ }\bibfield  {title} {\enquote {\bibinfo {title}
  {Floquet edge states with ultracold atoms},}\ }\href {\doibase
  10.1103/PhysRevA.89.063628} {\bibfield  {journal} {\bibinfo  {journal} {Phys.
  Rev. A}\ }\textbf {\bibinfo {volume} {89}},\ \bibinfo {pages} {063628}
  (\bibinfo {year} {2014})}\BibitemShut {NoStop}%
\bibitem [{\citenamefont {Maczewsky}\ \emph {et~al.}(2017)\citenamefont
  {Maczewsky}, \citenamefont {Zeuner}, \citenamefont {Nolte},\ and\
  \citenamefont {Szameit}}]{Maczewsky17_NatCommun}%
  \BibitemOpen
  \bibfield  {author} {\bibinfo {author} {\bibfnamefont {Lukas~J.}\
  \bibnamefont {Maczewsky}}, \bibinfo {author} {\bibfnamefont {Julia~M.}\
  \bibnamefont {Zeuner}}, \bibinfo {author} {\bibfnamefont {Stefan}\
  \bibnamefont {Nolte}}, \ and\ \bibinfo {author} {\bibfnamefont {Alexander}\
  \bibnamefont {Szameit}},\ }\bibfield  {title} {\enquote {\bibinfo {title}
  {Observation of photonic anomalous {F}loquet topological insulators},}\
  }\href {\doibase 10.1038/ncomms13756} {\bibfield  {journal} {\bibinfo
  {journal} {Nat. Commun.}\ }\textbf {\bibinfo {volume} {8}} (\bibinfo {year}
  {2017}),\ 10.1038/ncomms13756}\BibitemShut {NoStop}%
\bibitem [{\citenamefont {Mukherjee}\ \emph {et~al.}(2017)\citenamefont
  {Mukherjee}, \citenamefont {Spracklen}, \citenamefont {Valiente},
  \citenamefont {Andersson}, \citenamefont {Öhberg}, \citenamefont {Goldman},\
  and\ \citenamefont {Thomson}}]{Mukherjee17_NatComm}%
  \BibitemOpen
  \bibfield  {author} {\bibinfo {author} {\bibfnamefont {Sebabrata}\
  \bibnamefont {Mukherjee}}, \bibinfo {author} {\bibfnamefont {Alexander}\
  \bibnamefont {Spracklen}}, \bibinfo {author} {\bibfnamefont {Manuel}\
  \bibnamefont {Valiente}}, \bibinfo {author} {\bibfnamefont {Erika}\
  \bibnamefont {Andersson}}, \bibinfo {author} {\bibfnamefont {Patrik}\
  \bibnamefont {Öhberg}}, \bibinfo {author} {\bibfnamefont {Nathan}\
  \bibnamefont {Goldman}}, \ and\ \bibinfo {author} {\bibfnamefont {Robert~R}\
  \bibnamefont {Thomson}},\ }\bibfield  {title} {\enquote {\bibinfo {title}
  {Experimental observation of anomalous topological edge modes in a slowly
  driven photonic lattice},}\ }\href {\doibase 10.1038/ncomms13918} {\bibfield
  {journal} {\bibinfo  {journal} {Nat. Commun.}\ }\textbf {\bibinfo {volume}
  {8}} (\bibinfo {year} {2017}),\ 10.1038/ncomms13918}\BibitemShut {NoStop}%
\bibitem [{\citenamefont {Budich}\ \emph {et~al.}(2017)\citenamefont {Budich},
  \citenamefont {Hu},\ and\ \citenamefont {Zoller}}]{Budich17_PRL_helical}%
  \BibitemOpen
  \bibfield  {author} {\bibinfo {author} {\bibfnamefont {Jan~Carl}\
  \bibnamefont {Budich}}, \bibinfo {author} {\bibfnamefont {Ying}\ \bibnamefont
  {Hu}}, \ and\ \bibinfo {author} {\bibfnamefont {Peter}\ \bibnamefont
  {Zoller}},\ }\bibfield  {title} {\enquote {\bibinfo {title} {Helical floquet
  channels in 1d lattices},}\ }\href {\doibase 10.1103/PhysRevLett.118.105302}
  {\bibfield  {journal} {\bibinfo  {journal} {Phys. Rev. Lett.}\ }\textbf
  {\bibinfo {volume} {118}},\ \bibinfo {pages} {105302} (\bibinfo {year}
  {2017})}\BibitemShut {NoStop}%
\bibitem [{\citenamefont {Titum}\ \emph {et~al.}(2016)\citenamefont {Titum},
  \citenamefont {Berg}, \citenamefont {Rudner}, \citenamefont {Refael},\ and\
  \citenamefont {Lindner}}]{Titum16_PRX}%
  \BibitemOpen
  \bibfield  {author} {\bibinfo {author} {\bibfnamefont {Paraj}\ \bibnamefont
  {Titum}}, \bibinfo {author} {\bibfnamefont {Erez}\ \bibnamefont {Berg}},
  \bibinfo {author} {\bibfnamefont {Mark~S.}\ \bibnamefont {Rudner}}, \bibinfo
  {author} {\bibfnamefont {Gil}\ \bibnamefont {Refael}}, \ and\ \bibinfo
  {author} {\bibfnamefont {Netanel~H.}\ \bibnamefont {Lindner}},\ }\bibfield
  {title} {\enquote {\bibinfo {title} {Anomalous floquet-anderson insulator as
  a nonadiabatic quantized charge pump},}\ }\href {\doibase
  10.1103/PhysRevX.6.021013} {\bibfield  {journal} {\bibinfo  {journal} {Phys.
  Rev. X}\ }\textbf {\bibinfo {volume} {6}},\ \bibinfo {pages} {021013}
  (\bibinfo {year} {2016})}\BibitemShut {NoStop}%
\bibitem [{\citenamefont {Gross}\ and\ \citenamefont
  {Bakr}(2021)}]{GrossBakr21_NatPhys}%
  \BibitemOpen
  \bibfield  {author} {\bibinfo {author} {\bibfnamefont {C.}~\bibnamefont
  {Gross}}\ and\ \bibinfo {author} {\bibfnamefont {W.S.}\ \bibnamefont
  {Bakr}},\ }\bibfield  {title} {\enquote {\bibinfo {title} {Quantum gas
  microscopy for single atom and spin detection},}\ }\href
  {https://doi.org/10.1038/s41567-021-01370-5} {\bibfield  {journal} {\bibinfo
  {journal} {Nat. Phys.}\ }\textbf {\bibinfo {volume} {17}},\ \bibinfo {pages}
  {1316} (\bibinfo {year} {2021})},\ \bibinfo {note} {and references
  therein}\BibitemShut {NoStop}%
\bibitem [{\citenamefont {Quelle}\ \emph {et~al.}(2017)\citenamefont {Quelle},
  \citenamefont {Weitenberg}, \citenamefont {Sengstock},\ and\ \citenamefont
  {Smith}}]{Quelle17_NJP}%
  \BibitemOpen
  \bibfield  {author} {\bibinfo {author} {\bibfnamefont {A}~\bibnamefont
  {Quelle}}, \bibinfo {author} {\bibfnamefont {C}~\bibnamefont {Weitenberg}},
  \bibinfo {author} {\bibfnamefont {K}~\bibnamefont {Sengstock}}, \ and\
  \bibinfo {author} {\bibfnamefont {C~Morais}\ \bibnamefont {Smith}},\
  }\bibfield  {title} {\enquote {\bibinfo {title} {Driving protocol for a
  floquet topological phase without static counterpart},}\ }\href {\doibase
  10.1088/1367-2630/aa8646} {\bibfield  {journal} {\bibinfo  {journal} {New
  Journal of Physics}\ }\textbf {\bibinfo {volume} {19}},\ \bibinfo {pages}
  {113010} (\bibinfo {year} {2017})}\BibitemShut {NoStop}%
\bibitem [{\citenamefont {Nathan}\ and\ \citenamefont
  {Rudner}(2015)}]{nathan2015topological}%
  \BibitemOpen
  \bibfield  {author} {\bibinfo {author} {\bibfnamefont {Frederik}\
  \bibnamefont {Nathan}}\ and\ \bibinfo {author} {\bibfnamefont {Mark~S}\
  \bibnamefont {Rudner}},\ }\bibfield  {title} {\enquote {\bibinfo {title}
  {Topological singularities and the general classification of floquet--bloch
  systems},}\ }\href
  {https://iopscience.iop.org/article/10.1088/1367-2630/17/12/125014}
  {\bibfield  {journal} {\bibinfo  {journal} {New Journal of Physics}\ }\textbf
  {\bibinfo {volume} {17}},\ \bibinfo {pages} {125014} (\bibinfo {year}
  {2015})}\BibitemShut {NoStop}%
\bibitem [{See()}]{SeeSupplement}%
  \BibitemOpen
  \href@noop {} {\enquote {\bibinfo {title} {{S}ee {S}upplementary material at
  xxxx for details.}}\ }\BibitemShut {NoStop}%
\bibitem [{\citenamefont {Kaufman}\ and\ \citenamefont
  {Ni}(2021)}]{KaufmanNi21_NatPhys}%
  \BibitemOpen
  \bibfield  {author} {\bibinfo {author} {\bibfnamefont {A.M.}\ \bibnamefont
  {Kaufman}}\ and\ \bibinfo {author} {\bibfnamefont {KK.}\ \bibnamefont {Ni}},\
  }\bibfield  {title} {\enquote {\bibinfo {title} {Quantum science with optical
  tweezer arrays of ultracold atoms and molecules.}}\ }\href
  {https://doi.org/10.1038/s41567-021-01357-2} {\bibfield  {journal} {\bibinfo
  {journal} {Nat. Phys.}\ }\textbf {\bibinfo {volume} {17}},\ \bibinfo {pages}
  {1324} (\bibinfo {year} {2021})},\ \bibinfo {note} {and references
  therein.}\BibitemShut {Stop}%
\bibitem [{\citenamefont {Braun}\ \emph {et~al.}(2023)\citenamefont {Braun},
  \citenamefont {Saint-Jalm}, \citenamefont {Hesse}, \citenamefont {Arceri},
  \citenamefont {Bloch},\ and\ \citenamefont {Aidelsburger}}]{MunichExpWP}%
  \BibitemOpen
  \bibfield  {author} {\bibinfo {author} {\bibfnamefont {Christoph}\
  \bibnamefont {Braun}}, \bibinfo {author} {\bibfnamefont {Rapha\"{e}l}\
  \bibnamefont {Saint-Jalm}}, \bibinfo {author} {\bibfnamefont {Alexander}\
  \bibnamefont {Hesse}}, \bibinfo {author} {\bibfnamefont {Johannes}\
  \bibnamefont {Arceri}}, \bibinfo {author} {\bibfnamefont {Immanuel}\
  \bibnamefont {Bloch}}, \ and\ \bibinfo {author} {\bibfnamefont {Monika}\
  \bibnamefont {Aidelsburger}},\ }\href {\doibase 10.48550/arXiv.2304.01980}
  {\enquote {\bibinfo {title} {Real-space detection and manipulation of
  topological edge modes with ultracold atoms},}\ } (\bibinfo {year} {2023}),\
  \Eprint {http://arxiv.org/abs/2304.01980} {arXiv:2304.01980} \BibitemShut
  {NoStop}%
\bibitem [{\citenamefont {Nixon}\ \emph {et~al.}(2023)\citenamefont {Nixon},
  \citenamefont {\"{U}nal},\ and\ \citenamefont
  {Schneider}}]{Nixon23_arxiv_localdrive}%
  \BibitemOpen
  \bibfield  {author} {\bibinfo {author} {\bibfnamefont {Georgia~M}\
  \bibnamefont {Nixon}}, \bibinfo {author} {\bibfnamefont {F~Nur}\ \bibnamefont
  {\"{U}nal}}, \ and\ \bibinfo {author} {\bibfnamefont {Ulrich}\ \bibnamefont
  {Schneider}},\ }\bibfield  {title} {\enquote {\bibinfo {title} {Individually
  tunable tunnelling coefficients in optical lattices using local periodic
  driving},}\ }\href {\doibase 10.48550/arXiv.2309.12124} {\bibfield  {journal}
  {\bibinfo  {journal} {arXiv preprint arXiv:2309.12124}\ } (\bibinfo {year}
  {2023}),\ 10.48550/arXiv.2309.12124}\BibitemShut {NoStop}%
\bibitem [{\citenamefont {Haldane}(1988)}]{Haldane88_PRL}%
  \BibitemOpen
  \bibfield  {author} {\bibinfo {author} {\bibfnamefont {F.~D.~M.}\
  \bibnamefont {Haldane}},\ }\bibfield  {title} {\enquote {\bibinfo {title}
  {Model for a quantum hall effect without landau levels: Condensed-matter
  realization of the "parity anomaly"},}\ }\href {\doibase
  10.1103/PhysRevLett.61.2015} {\bibfield  {journal} {\bibinfo  {journal}
  {Phys. Rev. Lett.}\ }\textbf {\bibinfo {volume} {61}},\ \bibinfo {pages}
  {2015--2018} (\bibinfo {year} {1988})}\BibitemShut {NoStop}%
\bibitem [{\citenamefont {Zhang}\ \emph {et~al.}(2022)\citenamefont {Zhang},
  \citenamefont {Yi}, \citenamefont {Zhang}, \citenamefont {Jiao},
  \citenamefont {Shi}, \citenamefont {Y.}, \citenamefont {Z.}, \citenamefont
  {L.}, \citenamefont {Chen},\ and\ \citenamefont
  {Pan}}]{ZhangPan22_arXiv_floeuetBIS}%
  \BibitemOpen
  \bibfield  {author} {\bibinfo {author} {\bibfnamefont {J.-Y.}\ \bibnamefont
  {Zhang}}, \bibinfo {author} {\bibfnamefont {C.-R.}\ \bibnamefont {Yi}},
  \bibinfo {author} {\bibfnamefont {L.}~\bibnamefont {Zhang}}, \bibinfo
  {author} {\bibfnamefont {R.-H.}\ \bibnamefont {Jiao}}, \bibinfo {author}
  {\bibfnamefont {K.-Y.}\ \bibnamefont {Shi}}, \bibinfo {author} {\bibfnamefont
  {H.}~\bibnamefont {Y.}}, \bibinfo {author} {\bibfnamefont {W.}~\bibnamefont
  {Z.}}, \bibinfo {author} {\bibfnamefont {X.-J.}\ \bibnamefont {L.}}, \bibinfo
  {author} {\bibfnamefont {S.}~\bibnamefont {Chen}}, \ and\ \bibinfo {author}
  {\bibfnamefont {J.-W.}\ \bibnamefont {Pan}},\ }\href {\doibase
  10.48550/arXiv.2211.04739} {\enquote {\bibinfo {title} {Tuning anomalous
  floquet topological bands with ultracold atoms},}\ } (\bibinfo {year}
  {2022}),\ \Eprint {http://arxiv.org/abs/2211.04739} {arXiv:2211.04739}
  \BibitemShut {NoStop}%
\bibitem [{\citenamefont {Adiyatullin}\ \emph {et~al.}(2023)\citenamefont
  {Adiyatullin}, \citenamefont {Upreti}, \citenamefont {Lechevalier},
  \citenamefont {Evain}, \citenamefont {Copie}, \citenamefont {Suret},
  \citenamefont {Randoux}, \citenamefont {Delplace},\ and\ \citenamefont
  {Amo}}]{AdiyatullinAmo23_PRL}%
  \BibitemOpen
  \bibfield  {author} {\bibinfo {author} {\bibfnamefont {Albert~F.}\
  \bibnamefont {Adiyatullin}}, \bibinfo {author} {\bibfnamefont {Lavi~K.}\
  \bibnamefont {Upreti}}, \bibinfo {author} {\bibfnamefont {Corentin}\
  \bibnamefont {Lechevalier}}, \bibinfo {author} {\bibfnamefont {Clement}\
  \bibnamefont {Evain}}, \bibinfo {author} {\bibfnamefont {Francois}\
  \bibnamefont {Copie}}, \bibinfo {author} {\bibfnamefont {Pierre}\
  \bibnamefont {Suret}}, \bibinfo {author} {\bibfnamefont {Stephane}\
  \bibnamefont {Randoux}}, \bibinfo {author} {\bibfnamefont {Pierre}\
  \bibnamefont {Delplace}}, \ and\ \bibinfo {author} {\bibfnamefont {Alberto}\
  \bibnamefont {Amo}},\ }\bibfield  {title} {\enquote {\bibinfo {title}
  {Topological properties of floquet winding bands in a photonic lattice},}\
  }\href {\doibase 10.1103/PhysRevLett.130.056901} {\bibfield  {journal}
  {\bibinfo  {journal} {Phys. Rev. Lett.}\ }\textbf {\bibinfo {volume} {130}},\
  \bibinfo {pages} {056901} (\bibinfo {year} {2023})}\BibitemShut {NoStop}%
\end{thebibliography}%


\begin{thebibliography}{3}%
\makeatletter
\providecommand \@ifxundefined [1]{%
 \@ifx{#1\undefined}
}%
\providecommand \@ifnum [1]{%
 \ifnum #1\expandafter \@firstoftwo
 \else \expandafter \@secondoftwo
 \fi
}%
\providecommand \@ifx [1]{%
 \ifx #1\expandafter \@firstoftwo
 \else \expandafter \@secondoftwo
 \fi
}%
\providecommand \natexlab [1]{#1}%
\providecommand \enquote  [1]{``#1''}%
\providecommand \bibnamefont  [1]{#1}%
\providecommand \bibfnamefont [1]{#1}%
\providecommand \citenamefont [1]{#1}%
\providecommand \href@noop [0]{\@secondoftwo}%
\providecommand \href [0]{\begingroup \@sanitize@url \@href}%
\providecommand \@href[1]{\@@startlink{#1}\@@href}%
\providecommand \@@href[1]{\endgroup#1\@@endlink}%
\providecommand \@sanitize@url [0]{\catcode `\\12\catcode `\$12\catcode
  `\&12\catcode `\#12\catcode `\^12\catcode `\_12\catcode `\%12\relax}%
\providecommand \@@startlink[1]{}%
\providecommand \@@endlink[0]{}%
\providecommand \url  [0]{\begingroup\@sanitize@url \@url }%
\providecommand \@url [1]{\endgroup\@href {#1}{\urlprefix }}%
\providecommand \urlprefix  [0]{URL }%
\providecommand \Eprint [0]{\href }%
\providecommand \doibase [0]{http://dx.doi.org/}%
\providecommand \selectlanguage [0]{\@gobble}%
\providecommand \bibinfo  [0]{\@secondoftwo}%
\providecommand \bibfield  [0]{\@secondoftwo}%
\providecommand \translation [1]{[#1]}%
\providecommand \BibitemOpen [0]{}%
\providecommand \bibitemStop [0]{}%
\providecommand \bibitemNoStop [0]{.\EOS\space}%
\providecommand \EOS [0]{\spacefactor3000\relax}%
\providecommand \BibitemShut  [1]{\csname bibitem#1\endcsname}%
\let\auto@bib@innerbib\@empty
\bibitem [{\citenamefont {\"Unal}\ \emph {et~al.}(2019)\citenamefont {\"Unal},
  \citenamefont {Seradjeh},\ and\ \citenamefont
  {Eckardt}}]{Nur_PRL_HowToMeasure}%
  \BibitemOpen
  \bibfield  {author} {\bibinfo {author} {\bibfnamefont {F.~Nur}\ \bibnamefont
  {\"Unal}}, \bibinfo {author} {\bibfnamefont {Babak}\ \bibnamefont
  {Seradjeh}}, \ and\ \bibinfo {author} {\bibfnamefont {Andr\'e}\ \bibnamefont
  {Eckardt}},\ }\bibfield  {title} {\enquote {\bibinfo {title} {How to directly
  measure floquet topological invariants in optical lattices},}\ }\href
  {\doibase 10.1103/PhysRevLett.122.253601} {\bibfield  {journal} {\bibinfo
  {journal} {Phys. Rev. Lett.}\ }\textbf {\bibinfo {volume} {122}},\ \bibinfo
  {pages} {253601} (\bibinfo {year} {2019})}\BibitemShut {NoStop}%
\bibitem [{\citenamefont {Wintersperger}\ \emph {et~al.}(2020)\citenamefont
  {Wintersperger}, \citenamefont {Braun}, \citenamefont {{\"U}nal},
  \citenamefont {Eckardt}, \citenamefont {Liberto}, \citenamefont {Goldman},
  \citenamefont {Bloch},\ and\ \citenamefont
  {Aidelsburger}}]{wintersperger2020realization}%
  \BibitemOpen
  \bibfield  {author} {\bibinfo {author} {\bibfnamefont {Karen}\ \bibnamefont
  {Wintersperger}}, \bibinfo {author} {\bibfnamefont {Christoph}\ \bibnamefont
  {Braun}}, \bibinfo {author} {\bibfnamefont {F~Nur}\ \bibnamefont {{\"U}nal}},
  \bibinfo {author} {\bibfnamefont {Andr{\'e}}\ \bibnamefont {Eckardt}},
  \bibinfo {author} {\bibfnamefont {Marco~Di}\ \bibnamefont {Liberto}},
  \bibinfo {author} {\bibfnamefont {Nathan}\ \bibnamefont {Goldman}}, \bibinfo
  {author} {\bibfnamefont {Immanuel}\ \bibnamefont {Bloch}}, \ and\ \bibinfo
  {author} {\bibfnamefont {Monika}\ \bibnamefont {Aidelsburger}},\ }\bibfield
  {title} {\enquote {\bibinfo {title} {Realization of an anomalous floquet
  topological system with ultracold atoms},}\ }\href
  {https://www.nature.com/articles/s41567-020-0949-y#citeas} {\bibfield
  {journal} {\bibinfo  {journal} {Nature Physics}\ }\textbf {\bibinfo {volume}
  {16}},\ \bibinfo {pages} {1058--1063} (\bibinfo {year} {2020})}\BibitemShut
  {NoStop}%
\bibitem [{\citenamefont {Nathan}\ and\ \citenamefont
  {Rudner}(2015)}]{nathan2015topological}%
  \BibitemOpen
  \bibfield  {author} {\bibinfo {author} {\bibfnamefont {Frederik}\
  \bibnamefont {Nathan}}\ and\ \bibinfo {author} {\bibfnamefont {Mark~S}\
  \bibnamefont {Rudner}},\ }\bibfield  {title} {\enquote {\bibinfo {title}
  {Topological singularities and the general classification of floquet--bloch
  systems},}\ }\href
  {https://iopscience.iop.org/article/10.1088/1367-2630/17/12/125014}
  {\bibfield  {journal} {\bibinfo  {journal} {New Journal of Physics}\ }\textbf
  {\bibinfo {volume} {17}},\ \bibinfo {pages} {125014} (\bibinfo {year}
  {2015})}\BibitemShut {NoStop}%
\end{thebibliography}%

\end{document}


\title{ Supplementary Material for ``Wave packet dynamics and edge transport in anomalous Floquet topological phases" }
\author{Miguel F.~Mart\'{i}nez}
\affiliation{Department of Physics, KTH Royal Institute of Technology, 106 91, Stockholm, Sweden}
\affiliation{TCM Group, Cavendish Laboratory, University of Cambridge, JJ Thomson Avenue, Cambridge CB3 0HE, United Kingdom\looseness=-1}
\author{F.~Nur \"{U}nal}
\affiliation{TCM Group, Cavendish Laboratory, University of Cambridge, JJ Thomson Avenue, Cambridge CB3 0HE, United Kingdom\looseness=-1}

\maketitle
\setcounter{equation}{0}
\setcounter{figure}{0}
\setcounter{table}{0}
\setcounter{page}{1}
\makeatletter
\renewcommand{\theequation}{S\arabic{equation}}
\renewcommand{\thefigure}{S\arabic{figure}}

\onecolumngrid

\section{Calculation of the phase diagrams}
In this Supplementary section, we describe in further detail the process that we use to calculate the phase diagrams in Fig. 1, which has been introduced in Ref.~\cite{Nur_PRL_HowToMeasure} and successfully implemented in experiment to measure the winding numbers in anomalous Floquet topological phases in a gap specific way~\cite{wintersperger2020realization}.

The main idea of this method is to device a one-parameter family of drives that connects the target phase to a parameter region where the topological phase is known. Crucially, in the high-frequency regime where the Floquet replicas of quasienergy bands are well separated from each other, the topological classification can be captured by the equilibrium characterisation~\cite{Nur_PRL_HowToMeasure}. Namely, the winding number at the edge of the FBZ ($W_{\pi}$) vanishes and the Chern numbers become sufficient to classify the system, as can be also seen in analogy to phase bands~\cite{nathan2015topological}. Starting from this reference point (or any reference point in fact as long as the winding numbers are fully known), upon varying the parameter characterising the family of drives, the quasienergy bands change, eventually leading to topological transitions via band inversion. At each of the transitions between the initial and the final regions, one can compute a topological charge that dictates how the winding invariants in that gap change. Summing these charges together with the winding invariants characterising the initial phase yields the winding invariants of the target region~\cite{Nur_PRL_HowToMeasure}.

In this work, similarly, we consider the reference region of the family of drives to be the high-frequency limit, 
$\omega \gg \vert J \vert, \vert \Delta \vert$. In this limit, the $\pi$-gap becomes significantly larger than the $0$-gap, and remains trivial, making the static classification in terms of a Chern number applicable. 
We then vary the parameters $\omega, \Delta, \lambda$ towards the regions targeted in the phase diagrams.

At every topological transition point where a quasienergy gap closes, the topological charge can be calculated in the following way~\cite{Nur_PRL_HowToMeasure}. We first identify the parameters at which the singularity forms $\textbf{\textit{p}}_s\equiv (\textbf{\textit{k}}, \theta)$, where $\textbf{\textit{k}}$ is the quasimomentum and $\theta$ is the parameter defining the family of drives. The topological charge $q_s$ of the singularity in the gap $g$ is then defined as $q_s=\text{sgn}(\text{det}(S))$, where $S_{ij}=\partial h_{Fi}^{\theta}(\textbf{\textit{k}})/\partial p_j\vert_{\textbf{\textit{p}}_s}$. Here, $h_{Fi}^{\theta}(\textbf{\textit{k}})$ is the Floquet Hamiltonian $\mathcal{H}^\theta_F(\textbf{\textit{k}}) = \textbf{\textit{h}}^\theta_F (\textbf{\textit{k}}) \cdot \boldsymbol{\sigma}$ with the Pauli matrices $\boldsymbol{\sigma}$. Once we have calculated the charges related to all the topological transitions at both gaps, most importantly in a gap-specific way, the winding invariants of each gap can be calculated as: $W_g=W^0_g + \sum_s q_{s_g}$, with $[W^0_0, W^0_\pi]$ corresponding to the reference phase.%
%


\section{Wave packet dynamics}
In this Supplementary Section, we present further examples of some wave packet dynamics.

\begin{figure*}[h!]
     \includegraphics[width=1\linewidth]{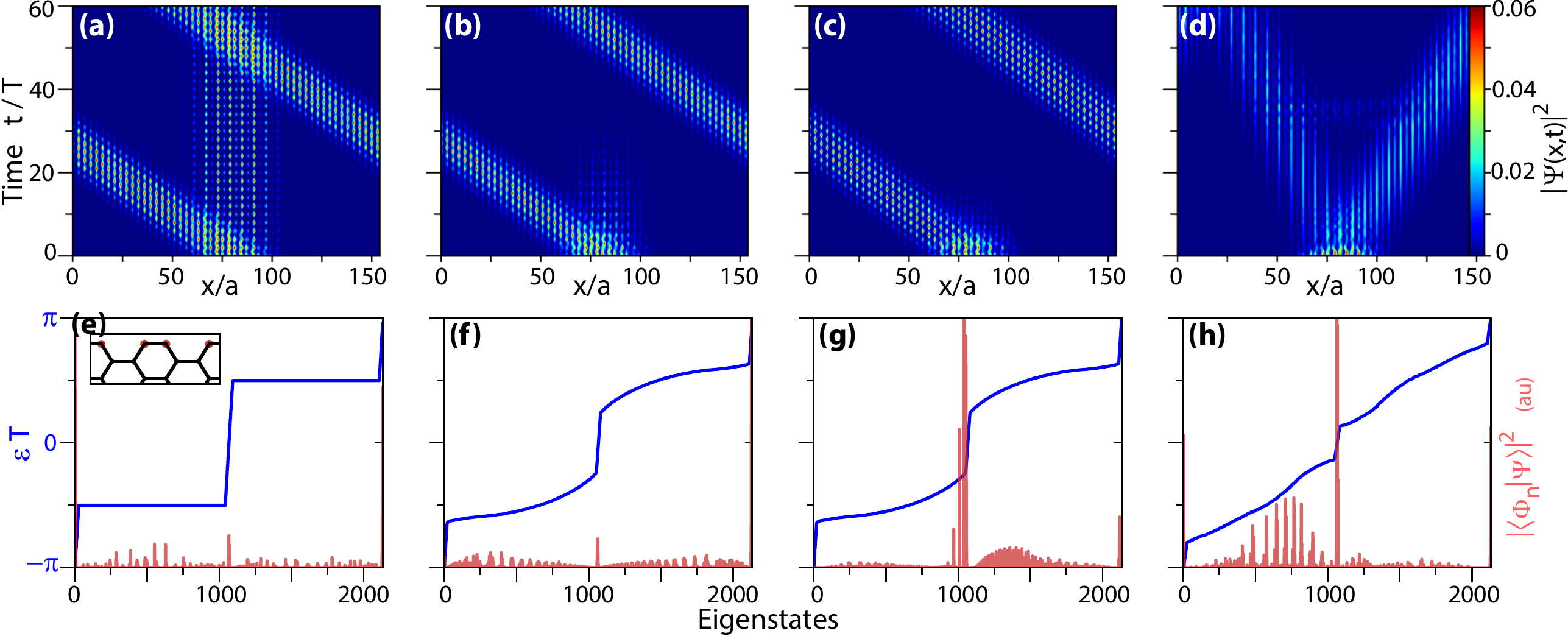}
    \caption{Initial overlap (bottom panel) of a wave packet with the Floquet eigenstates and its evolution (upper panel) at the edge sites (shaded two layers in the inset) for the different phases and parameters shown in Fig.~4 of the main text. In this case, the initial spread along the periodic direction ($x$) of the cylinder is $ \sigma_x=5$. Regardless of the phase, the dynamics become cleaner and do not show the characteristic branches developing in Fig.~4(c,d). This relates directly to the fact that the wave packet is less localised in real space, allowing to target more accurately the different gaps in momentum space, as revealed by the initial overlaps. }
    \label{fig:sup1}
\end{figure*}

Fig.~\ref{fig:sup1} demonstrates the same dynamics as in Fig.~4 of the main text for initial wave packets with larger $\sigma_x$ spread, i.e.~wider wave packets along the edge of the cylinder. In particular, compared to the more localised wave packets presented in the main text, the overlaps for $\sigma_x=5$ show naturally lesser spread in the momentum space. This results in clearer dynamics, since different branches travelling with different velocities that originate from the residual population of the other gap are absent. The population of the gap targeted by the kick (or by the absence of it) is enhanced, where the bulk contribution is also a suppressed. 

\begin{figure*}[b!]
  \centering\includegraphics[width=1\linewidth]{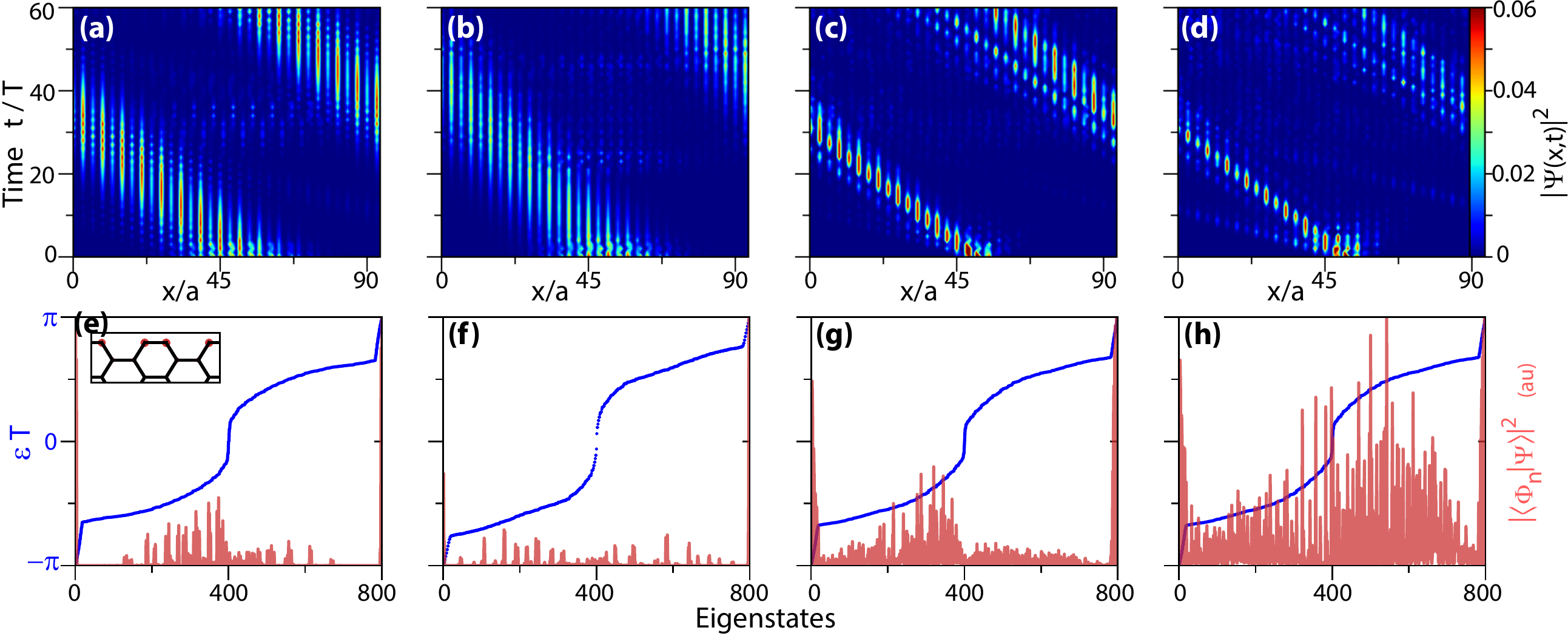}
    \caption{Initial overlap (bottom panel) of a wave packet with the Floquet eigenstates and its evolution (upper panel) at the edge sites (shaded two layers in the inset) in a cylinder with $N_x=104$, $N_y=41$ layers, and for the second driving protocol described on the text. (a, e) Unkicked wave packet in the $[1,1]$ phase for $\omega =7$, $\delta=0$, $\lambda=4.5$. (b, f) Unkicked $[0, 1]$ phase for $\omega =6.5$, $\delta=1$, $\lambda=3$. The comparison between the $[1,1]$ and $[0, 1]$ phases shows that we recover the same behaviour as in the first driving protocol, having more robust dynamics in the anomalous phase.  (c, g) Wave packet in the $[1,1]$ phase for $\omega =7$, $\delta=2$, $\lambda=3$ without a kick and (d, h) with $\textbf{\textit{q}}=(\pi/\sqrt{3}, 0)$. Even when the zero gap is very narrow, the kick allows us to populate it more strongly than when there is no kick, which can be appreciated by the development of a faint channel traveling with smaller velocity than the main channel carried by the $\pi$ gap. The initial spread of the different wave packets shown is given by $\sigma_x=5$ in the first two figures, and $\sigma_x=1$ in the last two. while $\sigma_y=0.5$ is kept fixed. }
    \label{fig:sup2}
\end{figure*} 

Moreover, as shown in the main text, the observed dynamics and structure are generic also for the second driving protocol. Complementing Fig.~5 of the main text, we here show examples of the dynamics in the $[1,1]$ anomalous phase and in the Haldane-like $[0,1]$ phase using the second driving protocol in Fig.~\ref{fig:sup2}(a) and (b) respectively. Similar to the the first driving protocol, we observe that the different edge modes contributing to the transport are overall more strongly populated in the anomalous phase than in the Haldane-like phase. Furthermore, to demonstrate the dynamics of the second drive upon applying initial kicks to the wave packet, we present the time evolution of a wave packet without and with a kick in Fig.~\ref{fig:sup2}(c) and (d) respectively, wave packet in the $[1,1]$ phase. In general, we obtain that the zero gap is narrower than the pi gap, and this can be seen in the overlaps as regardless of the kick the $\pi$-gap is strongly populated. However, with a kick, the population of the zero gap is enhanced compared to the absence of a kick, and its transport channel becomes more pronounced in the dynamics. Hence, the kick still allows to target the different gaps as expected.

\begin{figure}
  \centering\includegraphics[width=0.4\linewidth]{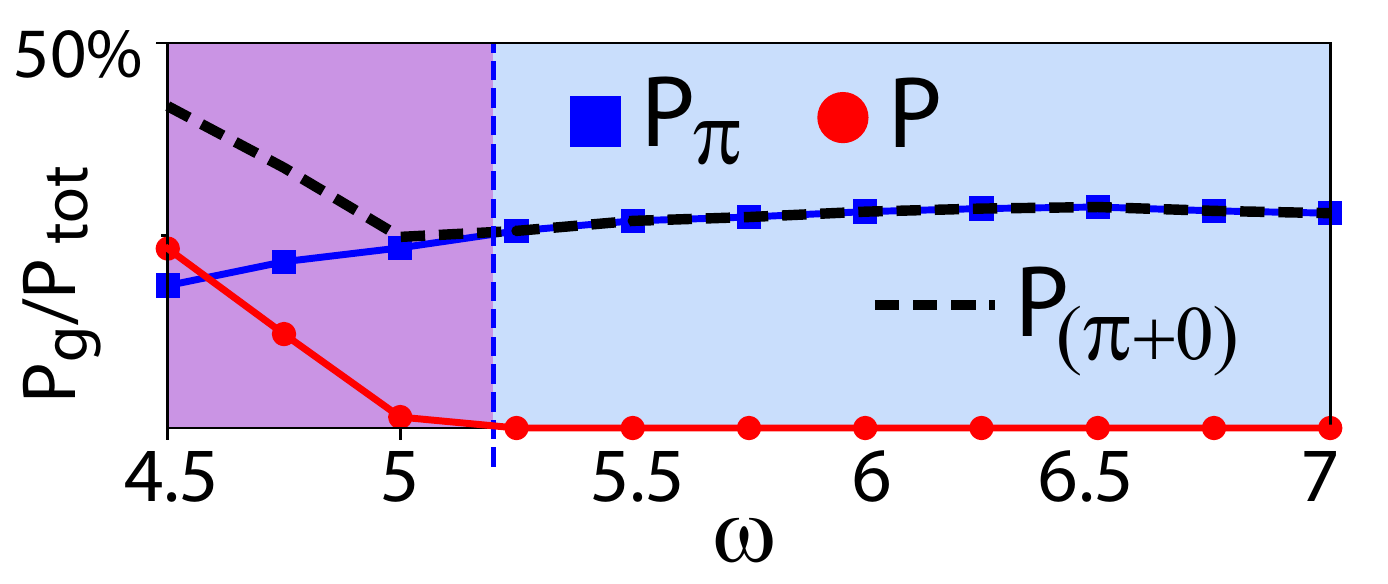}
     \caption{ Percentage of the total probability carried by the edge states in each gap, along the $\lambda=3$ cut of Fig.~1b of the main text, which crosses different phases at $\delta=2$. The total edge state population is overall higher in the $[-1,1]$ (purple shaded area) than in the $[0,1]$ phase (blue shaded area), for a wave packet without a kick. This indicates that the edge transport is overall more robust in the anomalous phase.}
    \label{fig:sup3}
\end{figure}

\newpage
Similar to the Fig.~5, we initialise a wave packet at the edge of a cylinder across a cut in the phase diagram given in Fig.~1b for the second driving protocol, crossing the anomalous $[-1,1]$ phase and the Haldane-like phase. In Fig.\ref{fig:sup3}, we demonstrate the total probability carried by the edge modes, which is overall higher in the anomalous phase than the single edge channel present in the Haldane-like phase, despite the opposite chirailities of the edge modes in the former.

\bibliography{references}